\documentclass[preprint2]{aastex6}

\shorttitle{Solar-Like Oscillations of $\theta$ Cyg}
\shortauthors{Guzik et al.}

\begin{document}


\title{Detection of Solar-Like Oscillations, Observational Constraints, and Stellar Models for $\theta$ Cyg, the Brightest Star Observed by the {\it Kepler} Mission}


\author{J. A. Guzik\altaffilmark{1}, G. Houdek\altaffilmark{2}, W. J. Chaplin\altaffilmark{3,2}, B. Smalley\altaffilmark{4}, D. W. Kurtz\altaffilmark{5},  R.L. Gilliland\altaffilmark{6}, F. Mullally\altaffilmark{7}, J.F. Rowe\altaffilmark{7}, S. T. Bryson\altaffilmark{8}, M. D. Still\altaffilmark{8,9}, V. Antoci\altaffilmark{2}, T. Appourchaux\altaffilmark{10},  S. Basu\altaffilmark{11}, T. R. Bedding\altaffilmark{12,2}, O. Benomar\altaffilmark{12,13}, R. A. Garcia\altaffilmark{14}, D. Huber\altaffilmark{12,2}, H. Kjeldsen\altaffilmark{2}, D. W. Latham\altaffilmark{15},  T.S. Metcalfe\altaffilmark{16}, P.~I.~P\'{a}pics\altaffilmark{17}, T. R. White\altaffilmark{12,18,2}, C. Aerts\altaffilmark{17,19}, J. Ballot\altaffilmark{34}, T. S. Boyajian\altaffilmark{11}, M. Briquet\altaffilmark{20}, H. Bruntt\altaffilmark{2,21},  L. A. Buchhave\altaffilmark{22,23}, T. L. Campante\altaffilmark{3,2}, G. Catanzaro\altaffilmark{24}, J. Christensen-Dalsgaard\altaffilmark{2}, G. R. Davies\altaffilmark{3,2,14}, G. Do\u{g}an\altaffilmark{2,26,27}, D. Dragomir\altaffilmark{28}, A. P. Doyle\altaffilmark{25,4},Y. Elsworth\altaffilmark{3,2}, A. Frasca\altaffilmark{24},  P. Gaulme\altaffilmark{29, 30}, M. Gruberbauer\altaffilmark{31}, R. Handberg\altaffilmark{2}, S. Hekker\altaffilmark{32,2},  C. Karoff\altaffilmark{2,26}, H. Lehmann\altaffilmark{33}, P. Mathias\altaffilmark{34,35}, S. Mathur\altaffilmark{16}, A. Miglio\altaffilmark{3,2}, J. Molenda-\.Zakowicz\altaffilmark{36}, B. Mosser\altaffilmark{37}, S. J. Murphy\altaffilmark{12,2}, C. R\'egulo\altaffilmark{38,39}, V. Ripepi\altaffilmark{40}, D. Salabert\altaffilmark{14}, S. G. Sousa\altaffilmark{41}, D. Stello\altaffilmark{12,2}, K. Uytterhoeven\altaffilmark{38,39}}

\altaffiltext{1}{Los Alamos National Laboratory, XTD-NTA, MS T-082, Los Alamos, NM  87545 USA}
\altaffiltext{2}{Stellar Astrophysics Centre, Department of Physics and Astronomy, Aarhus University, Ny Munkegade 120, DK-8000 Aarhus C, Denmark}
\altaffiltext{3}{School of Physics and Astronomy, University of Birmingham, Birmingham B15 2TT, UK}
\altaffiltext{4}{Astrophysics Group, School of Physical \& Geographical Sciences, Lennard-Jones Laboratories, Keele University, Staffordshire, ST5 5BG, UK}
\altaffiltext{5}{Jeremiah Horrocks Institute, University of Central Lancashire, Preston PR1 2HE, UK}
\altaffiltext{6}{Center for Exoplanets and Habitable Worlds, The Pennsylvania State University, University Park, PA  16802 USA}
\altaffiltext{7}{SETI Institute/NASA Ames Research Center, Moffett Field, CA 94035 USA}
\altaffiltext{8}{NASA Ames Research Center, Bldg. 244, MS-244-30, Moffett Field, CA  94035 USA}
\altaffiltext{9}{Bay Area Environmental Research Institute, 560 Third Street W., Sonoma, CA 95476 USA}
\altaffiltext{10}{Institut d'Astrophysique Spatiale, Universit\`e de Paris Sud--CNRS, Batiment 121, F-91405 ORSAY Cedex, France}
\altaffiltext{11}{Department of Astronomy, Yale University, PO Box 208101, New Haven, CT 06520-8101 USA}
\altaffiltext{12}{Sydney Institute for Astronomy (SIfA), School of Physics, University of Sydney, NSW 2006, Australia}
\altaffiltext{13}{NYUAD Institute, Center for Space Science, New York University Abu Dhabi, PO Box 129188, Abu Dhabi, UAE}
\altaffiltext{14}{Laboratoire AIM, CEA/DRF -- CNRS - Univ. Paris Diderot -- IRFU/SAp, Centre de Saclay, 91191 Gif-sur-Yvette Cedex, France}
\altaffiltext{15}{Harvard-Smithsonian Center for Astrophysics, 60 Garden Street, Cambridge, MA 02138  USA}
\altaffiltext{16}{Space Science Institute, 4750 Walnut Street, Suite 205, Boulder, CO 80301 USA}
\altaffiltext{17}{Instituut voor Sterrenkunde, KU Leuven, Celestijnenlaan 200D, B-3001 Leuven, Belgium}
\altaffiltext{18}{Australian Astronomical Observatory, PO Box 915, North Ryde, NSW 1670, Australia}
\altaffiltext{19}{Department of Astrophysics/IMAPP, Radboud University Nijmegen, 6500 GL Nijmegen, The Netherlands}
\altaffiltext{20}{Institut d'Astrophysique et de G\'eophysique, Universit\'e de Li\`ege, Quartier Agora, All\'ee du 6 ao\^ut 19C, B-4000, Li\`ege, Belgium}
\altaffiltext{21}{Aarhus Katedralskole, Skolegyde 1, DK-8000 Aarhus C, Denmark}
\altaffiltext{22}{Niels Bohr Institute, University of Copenhagen, DK-2100 Copenhagen, Denmark}
\altaffiltext{23}{Centre for Star and Planet Formation, Natural History Museum of Denmark, University of Copenhagen, DK-1350 Copenhagen, Denmark}
\altaffiltext{24}{INAF-Osservatorio Astrofisico di Catania, Via S.Sofia 78, I-95123 Catania, Italy}
\altaffiltext{25}{Department of Physics, University of Warwick, Gibbet Hill Road, Coventry CV4 7AL, UK}
\altaffiltext{26}{Department of Geoscience, Aarhus University, Hoegh-Guldbergs Gade 2, DK-8000, Aarhus C, Denmark}
\altaffiltext{27}{High Altitude Observatory, National Center for Atmospheric Research, PO Box 3000, Boulder, CO 80307, USA}
\altaffiltext{28}{The Department of Astronomy and Astrophysics, University of Chicago, 5640 S Ellis Ave, Chicago, IL 60637, USA}
\altaffiltext{29}{Apache Point Observatory, Sloan Digital Sky Survey, PO Box 59, Sunspot, NM 88349, USA}
\altaffiltext{30}{New Mexico State University, Department of Astronomy, PO Box 30001, Las Cruces, NM 88003-4500, USA}
\altaffiltext{31}{Institute for Computational Astrophysics, Department of Astronomy and Physics, Saint Mary's University, Halifax, NS B3H 3C3, Canada}
\altaffiltext{32}{Max Planck Institute for Solar System Research, SAGE research group, Justus-von-Liebig-Weg 3, 37077 Göttingen, Germany}
\altaffiltext{33}{Th\"{u}ringer Landessternwarte Tautenburg (TLS), Sternwarte 5, D-07778 Tautenburg, Germany}
\altaffiltext{34}{Universit\'e de Toulouse, UPS-OMP, IRAP, 65000, Tarbes, France}
\altaffiltext{35}{CNRS, IRAP, 57 avenue d'Azereix, BP 826, 65008, Tarbes, France}
\altaffiltext{36}{Instytut Astronomiczny Uniwersytetu Wroc{\l}awskiego, ul. Kopernika 11, 51-622 Wroc{\l}aw, Poland}
\altaffiltext{37}{LESIA--Observatoire de Paris/CNRS, Sorbonne Universit\'es, UMC Univ. Paris 06, Univ. Paris Diderot, Sorbonne Paris Cit\'e, France}
\altaffiltext{38}{Instituto de Astrof\'isica de Canarias, 38205, La Laguna, Tenerife, Spain}
\altaffiltext{39}{Universidad de La Laguna, Dpto de Astrof\'isica, 38206, Tenerife, Spain}
\altaffiltext{40}{INAF-Osservatorio Astronomico di Capodimonte, Via Moiariello 16, I-80131 Napoli, Italy}
\altaffiltext{41}{Instituto de Astrof\'isica e Ci\^encias do Espa\c{c}o, Universidade do Porto, CAUP, Rua das Estrelas, 4150-762 Porto, Portugal}

\begin{abstract}
$\theta$~Cygni  is an F3 spectral-type main-sequence star with visual magnitude V=4.48.  This star was the brightest star observed by the original {\it Kepler} spacecraft mission.  Short-cadence (58.8 s) photometric data using a custom aperture were obtained during Quarter 6 (June-September 2010) and subsequently in Quarters 8 and 12-17.  We present analyses of the solar-like oscillations based on Q6 and Q8 data, identifying angular degree $l$ = 0, 1, and 2 oscillations in the range 1000-2700 $\mu$Hz, with a large frequency separation of 83.9 $\pm$ 0.4 $\mu$Hz, and frequency with maximum amplitude $\nu_{\rm max}$ = 1829 $\pm$ 54 $\mu$Hz.  We also present analyses of new ground-based spectroscopic observations, which, when combined with angular diameter measurements from interferometry and Hipparcos parallax, give $T_\mathrm{eff}$ = 6697 $\pm$ 78 K, radius 1.49 $\pm$ 0.03 R$_{\odot}$, [Fe/H] = -0.02 $\pm$ 0.06 dex, log $g$ = 4.23 $\pm$ 0.03.  We calculate stellar models matching the constraints using several methods, including using the Yale Rotating Evolution Code and the Asteroseismic Modeling Portal.  The best-fit models have masses 1.35--1.39 M$_{\odot}$ and ages 1.0--1.6 Gyr.  $\theta$~Cyg's $T_\mathrm{eff}$ and log $g$ place it cooler than the red edge of the $\gamma$ Doradus instability region established from pre-{\it Kepler} ground-based observations, but just at the red edge derived from pulsation modeling.  The best-fitting models have envelope convection-zone base temperature of $\sim$320,000 to 395,000 K.  The pulsation models show $\gamma$ Dor gravity-mode pulsations driven by the convective-blocking mechanism, with periods of 0.3 to 1 day (frequencies 11 to 33 $\mu$Hz).  However, gravity modes were not detected in the {\it Kepler} data; one signal at 1.776 c d$^{-1}$ (20.56 $\mu$Hz) may be attributable to a faint, possibly background, binary.  Asteroseismic studies of $\theta$~Cyg, in conjunction with those for other A-F stars observed by {\it Kepler} and CoRoT, will help to improve stellar model physics to sort out the confusing relationship between $\delta$ Sct and $\gamma$ Dor pulsations and their hybrids, and to test pulsation driving mechanisms.


\end{abstract}





\section{Introduction}
\label{sec:introduction}

The mission of the NASA {\it Kepler} spacecraft, launched 2009 March 7, was to 
search for Earth-sized planets around Sun-like stars in a fixed field of view in the Cygnus-Lyra region using high-precision CCD 
photometry to detect planetary transits \citep{Borucki2010}. 
As a secondary mission {\it Kepler} surveyed and monitored over 10,000 stars for 
asteroseismology, using the intrinsic brightness variations caused by 
pulsations to infer the star's mass, age, and interior structure \citep{2010PASP..122..131G}.
After the failure of the second of four reaction wheels, the {\it Kepler} mission transitioned into a new phase, K2 \citep{2014PASP..126..398H}, observing fields near the ecliptic plane for about 90 days each, with a variety of science objectives including planet searches.

The $V\,=\,4.48$ F3 spectral-type main-sequence star $\theta$\,Cyg, also known as 13\,Cyg, HR\,7469, HD\,185395, 2MASS 19362654+5013155, HIP 96441, and KIC\,11918630, where KIC = {\it Kepler} Input Catalog \citep{2011AJ....142..112B}, is the brightest star that fell on 
active pixels in the original {\it Kepler} field of view.  $\theta$~Cyg is nearby 
and bright, so that high-precision ground-based data can be combined with high 
signal-to-noise and long time-series {\it Kepler} photometry to provide constraints 
for asteroseismology.  The position of $\theta$~Cyg in the HR~diagram is near that of known
$\gamma$\,Dor pulsators, suggesting the possibility that it may exhibit 
high-order gravity mode pulsations, which would probe the stellar interior just 
outside its convective core.  $\theta$~Cyg is also cool enough to exhibit solar-like
$p$-mode (acoustic) oscillations, which probe both the interior and envelope structure.

\begin{figure}
\center{\includegraphics[width=\columnwidth]{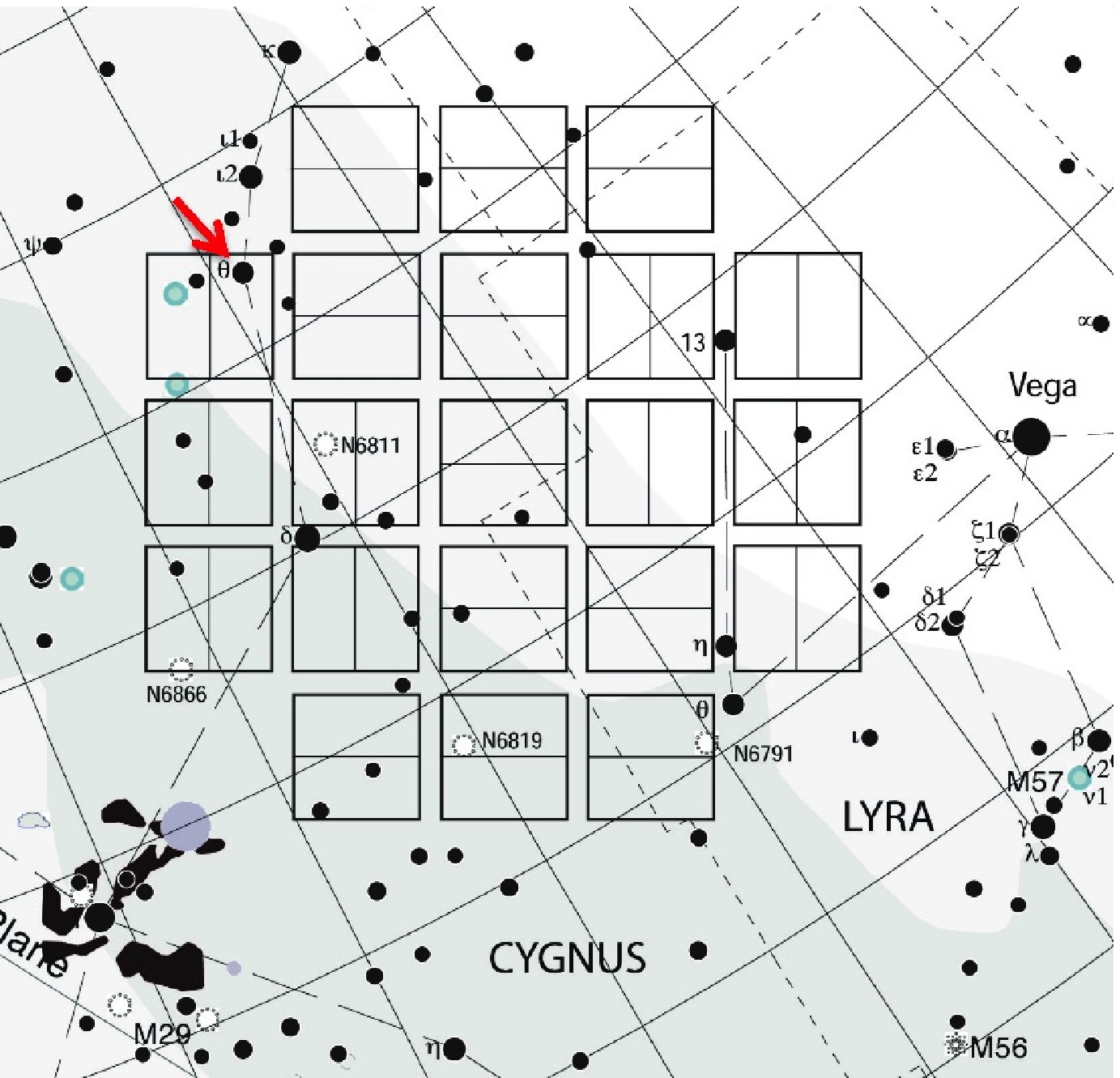}}
\caption{{\it Kepler} field of view with stars marked according to stellar magnitude created using ``The Sky'' astronomy software (http://www.bisque.com/sc/pages/TheSkyX-Editions.aspx).  The location of $\theta$~Cyg is shown by the red arrow, and is marked with the symbol $\theta$ and a filled circle designating magnitude 4-5.  Note that all stars brighter than $\theta$ Cyg fall in the regions between the CCD arrays to avoid saturating pixels.} 
\label{KeplerFOV}
\end{figure}



$\theta$ Cyg has been observed using adaptive optics \citep{2009A&A...506.1469D}.  It has a resolved binary M-dwarf companion of $\sim$0.35\,M$_{\odot}$ with separation 46\,AU.  Following the orbit for nearly an 
orbital period (unfortunately $\sim$~230 y) will eventually give an accurate 
dynamical mass for $\theta$ Cyg.  Also, the system shows a 150-d quasi-period in 
radial velocity, suggesting that one or more planets could accompany the 
stars \citep{2009A&A...506.1469D}. 

$\theta$ Cyg has also been observed using optical interferometry \citep[][see Section \ref{sec:interferometry}]{ 2012A&A...545A...5L, 2012ApJ...746..101B, 2013MNRAS.tmp.1445W}.  These observations provide tight constraints on the radius of  $\theta$ Cyg and therefore a very useful constraint for asteroseismology.

$\theta$ Cyg's projected rotational velocity is low; $v~\sin i$ = $3.4 \pm 0.4$ \,km s$^{-1}$ \citep[][see Section \ref{sec:spectroscopy}]{1984ApJ...281..719G}.  
If $\sin i$ is not too small, $\theta$ Cyg's slow rotation should simplify mode identification and pulsation modeling, as spherical approximations  and low-order perturbation theory for the rotational splitting should be adequate.

This paper is intended to provide background on the $\theta$~Cyg system and to be a first look at the {\it Kepler} photometry data and consequences for stellar models and asteroseismology.  We present light curves and detection of the solar-like $p$-modes based on {\it Kepler} data taken in observing Quarters 6 and 8 (Section \ref{sec:detection}).  We summarize ground-based observational constraints from the literature (Appendix A) and present analyses based on new spectroscopic observations (Section \ref{sec:spectroscopy}) and optical interferometry (Section \ref{sec:interferometry}).  We discuss inference of stellar parameters based on the large separation and frequency of maximum amplitude (Section \ref{sec:stellarparameters}), line widths (Section \ref{sec:damping}), and mode identification (Section \ref{sec:ModeID}).  We use the observed $p$-mode oscillation frequencies and mode identifications as constraints for stellar models using several methods (Section \ref{sec:models}).  We discuss predictions for $\gamma$ Dor $g$-mode pulsations (Section \ref{sec:gmodepredictions}), and results of a search for low frequencies consistent with $g$ modes (Section \ref{sec:gmodesearch}).  We conclude with motivation for continued study of $\theta$ Cyg (Section \ref{sec:conclusions}).

We do not include in this paper the analyses of data from Quarters 12-17 for several reasons.  First, we completed the bulk of this paper, including the spectroscopic analyses, and first asteroseismic analyses at the time when only the Q6 and Q8 data were available.  Second, a problem has emerged with the {\it Kepler} data reduction pipeline for the latest data release for short-cadence data\footnote{https://archive.stsci.edu/kepler/KSCI-19080-002.pdf} that will not be corrected until later in 2016; while $\theta$ Cyg is not on the list of affected stars, because $\theta$ Cyg required so many pixels and special processing, more work is needed to confirm that the problem has not introduced additional noise in the light curve.  We estimate that inclusion of the full time-series data will result in finding a few more frequencies, and will improve the precision of the frequencies obtained by a factor of $\sim$1.8.
Comparison of studies of the bright (V = 5.98) {\it Kepler} targets 16 Cyg A and B using one month versus thirty months of data show that the longer time series improved the accuracy and precision of results, but did not significantly change the frequencies or inferred stellar model parameters \citep{2012ApJ...748L..10M,2015ApJ...811L..37M}.

Detailed analyses making use of the remaining time-series data and the {\it Kepler} pixel data will be the subject of future papers.

\section{Detection of $\theta$ Cyg Solar-Like Oscillations by {\it Kepler}}
\label{sec:detection}

$\theta$\,Cyg is seven magnitudes brighter than the saturation limit of the {\it  Kepler} photometry.  Figure~\ref{KeplerFOV} shows the {\it Kepler} field of view superimposed on the constellations Cygnus and Lyra with $\theta$ Cyg on the CCD module at the top of the leftmost column in this figure.  {\it Kepler} stars are observed 
using masks that define the pixels to be stored for that star.  Special apertures can be defined to better conform to the distribution of charge for extremely saturated targets \citep[see, e.g.,][]{2011MNRAS.411..878K}. For $\theta$\,Cyg the number of recorded pixels required was reduced from  
$>$10,000 to $\sim$1,800 by using an improved special aperture.

$\theta$\,Cyg was observed 2010 June$-$September ({\it Kepler} Quarter 6) and 2011 Jan$-$March (Quarter 8) in short cadence (58.8 s integration; see \citet{2010ApJ...713L.160G} for details).  {\it Kepler} measurements were organized in quarters because the satellite performed a roll every three months to maintain the solar panels directed towards the Sun and the radiators to cool the focal plane in shadow.  Moreover, every month the satellite stopped data acquisition for less than 24 hours and pointed towards the Earth to transmit the stored data. Therefore, monthly interruptions occured in the {\it Kepler} observations.  More details on the {\it Kepler} window function can be found in \citet{2014A&A...568A..10G}.

\begin{figure*}
\mbox{
\includegraphics[scale=0.5]{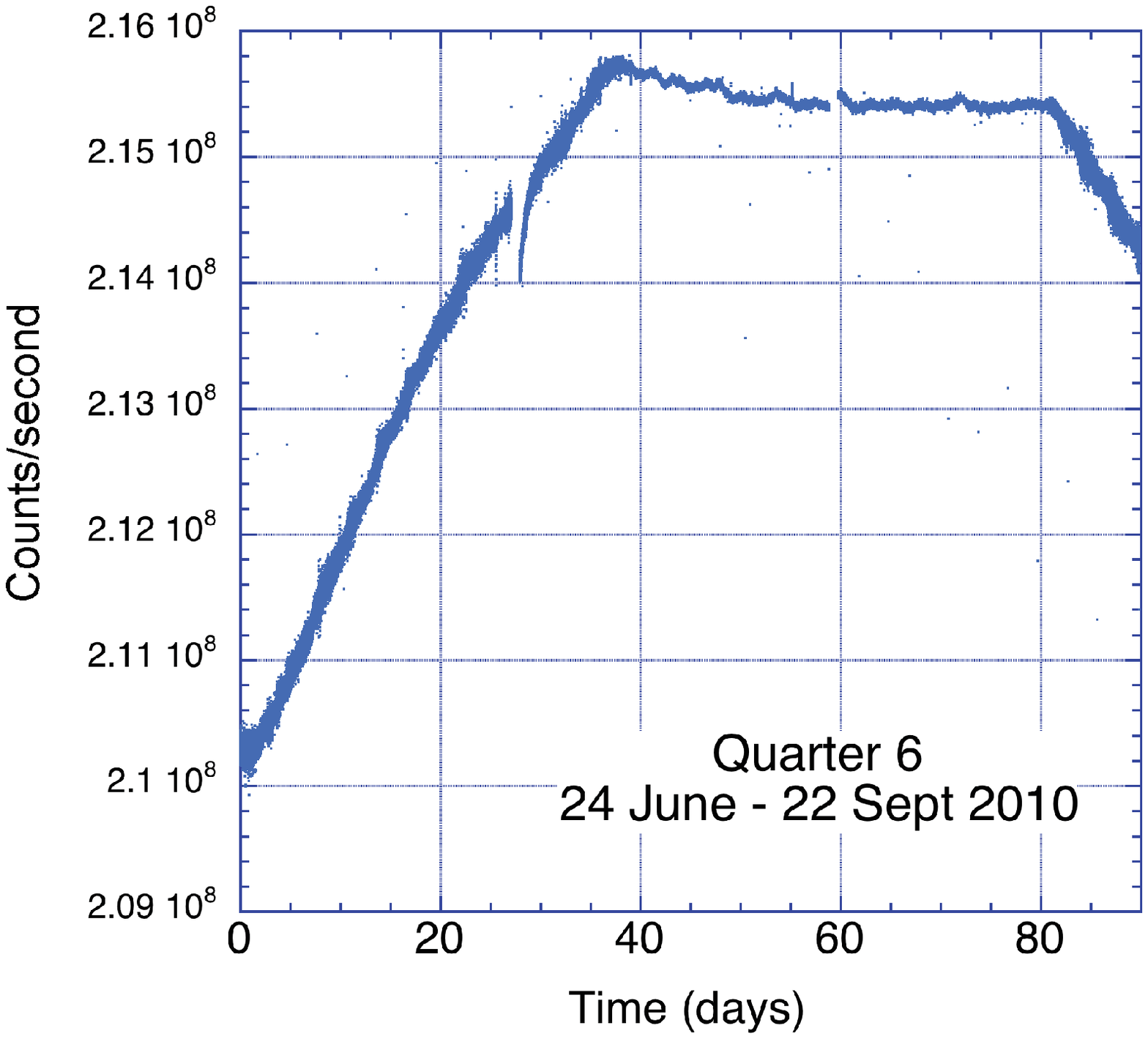}
\includegraphics[scale=0.5]{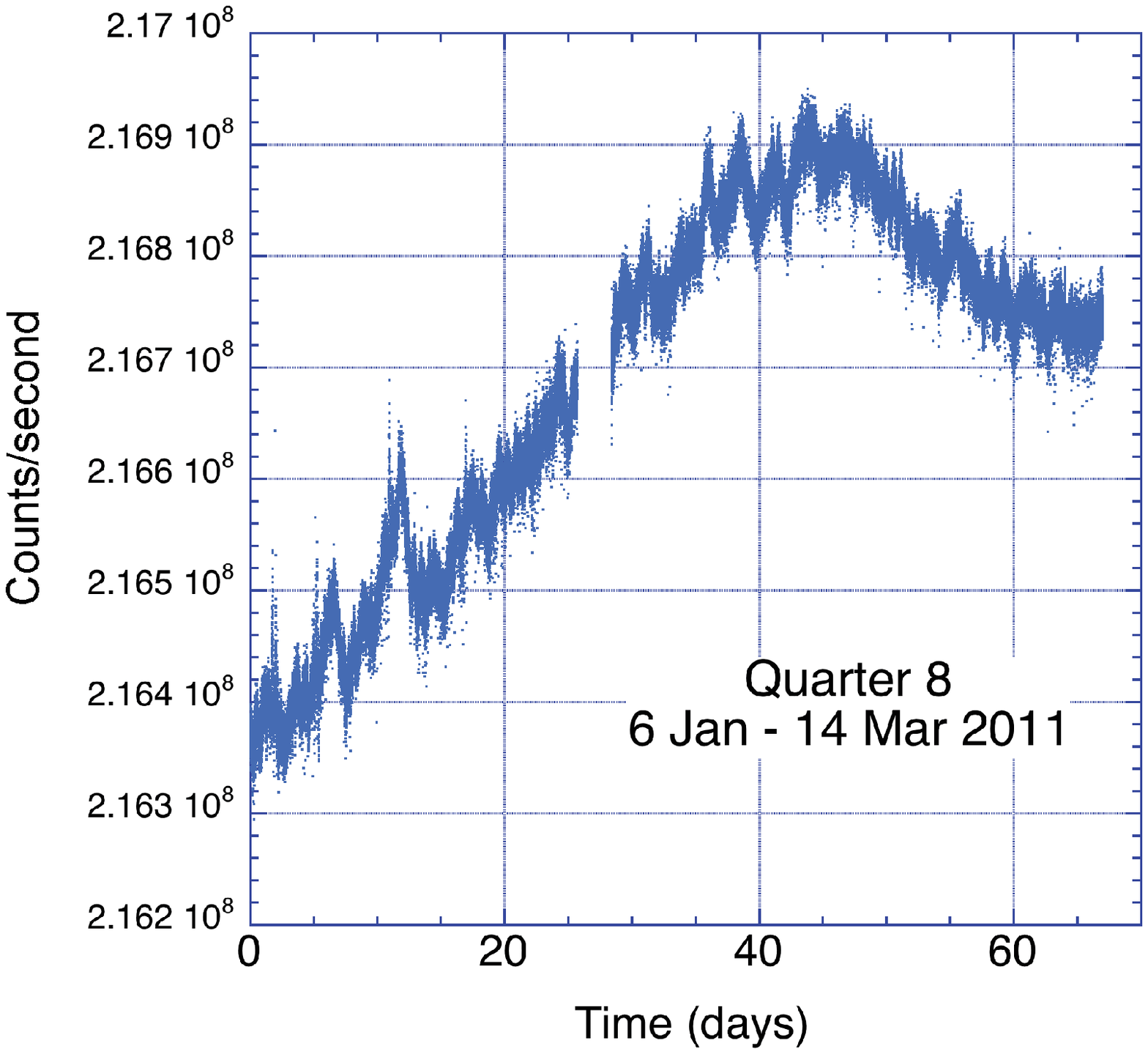}
}
\caption{{\it Kepler}  $\theta$\,Cyg unprocessed light curve for Quarter 6 (left) and Quarter 8 (right).  The custom 
aperture captured the target completely in Q6 only during 42 d (flat portion of curve) used in this analysis.  The spacecraft entered a safe mode for part of Q8, so only 67 days of data were obtained.} 
\label{rawlightcurves}
\end{figure*}

Figure~\ref{rawlightcurves} shows the 90-day
minimally processed short-cadence light curves for Q6 and Q8.
$\theta$\,Cyg was not well-captured by the dedicated mask for $\sim$50\% of Quarter 6 (a 
problem resolved for observations in subsequent quarters), so 42\,d of the best-quality data in the flat portion of the Q6 light curve were used in the 
pulsation analysis.  During Q8, the spacecraft entered a safe mode Dec. 22-Jan. 6, causing data loss at the beginning of the quarter, so only 67 d of data were obtained. 

\begin{figure}
\center{\includegraphics[width=\columnwidth]{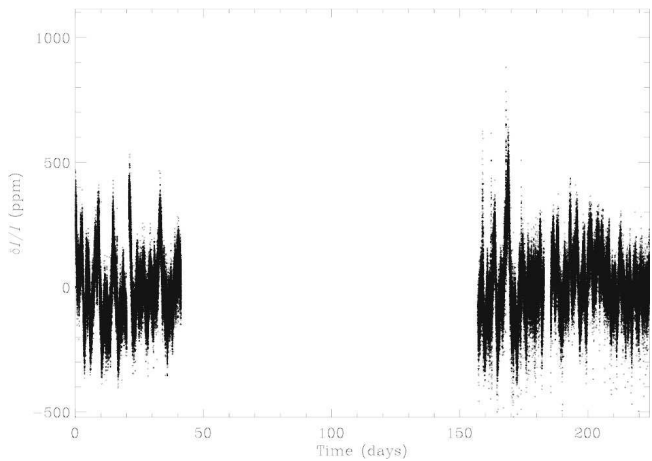}}
\caption{Combined Q6 and Q8 light curve after detrending and applying a 10-day triangular filter to remove low-frequency variations.} 
\label{Q6Q8LightCurve}
\end{figure}

These light curves were processed following the methods described by \citet{2011MNRAS.414L...6G} to remove outliers, jumps and drifts, as was done for other solar-like stars \citep[e.g.][]{2011A&A...534A...6C,2011ApJ...733...95M,2012A&A...543A..54A}, including the binary system 16 Cyg \citep{2012ApJ...748L..10M}, where a special treatment was also applied because it is composed of two very bright stars.  For the solar-like oscillation analysis of $\theta$ Cyg, we have removed the drifts by using a triangular smoothing filter with a width of 10 days (frequency 1.16 $\mu$Hz). The triangular smoothing filter is a rectangular (box car) filter of 10 days applied twice to the data; hence it is the convolution of two box cars, which is a triangle.  Figure~\ref{Q6Q8LightCurve} shows the resultant light curve for the Q6 and Q8 data.

\begin{figure*}
\center{\includegraphics[width=\textwidth]{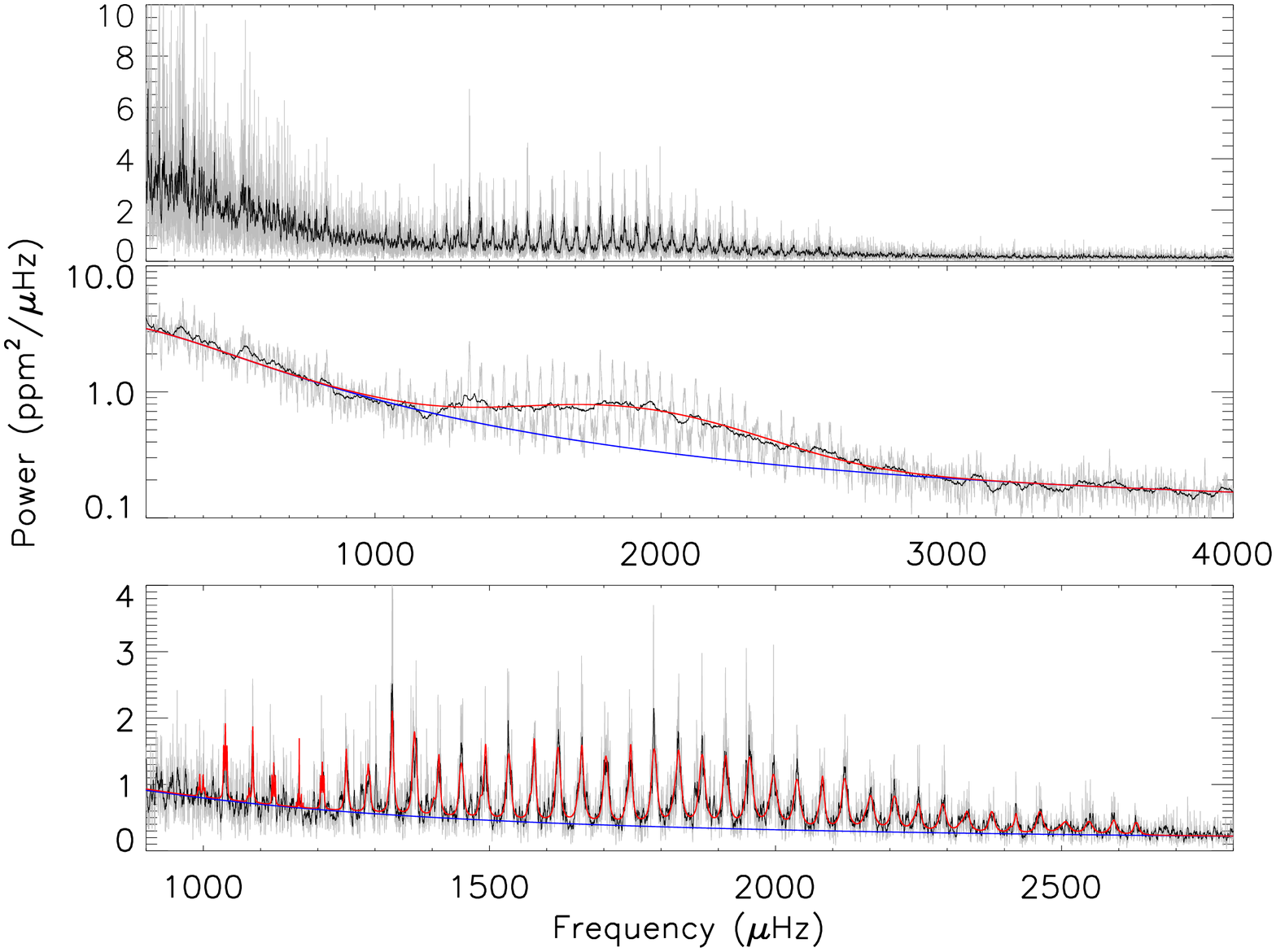}}
\caption{Top:  Power-density spectrum of the Q6 and Q8 data shown using the
minimally processed data (grey) and using a box-car smoothing of width $3$ $\mu$Hz (black).
Middle:  Power spectrum with a smoothing of width $3$ $\mu$Hz (grey) or
$\Delta\nu/2 = 42$ $\mu$Hz (black). Superimposed is shown the best fit of the
mode envelope with a Gaussian (red) and of the noise background (blue).
Bottom: Zoom-in on the modes. The power spectrum is smoothed over $0.5$ $\mu$Hz (grey) or $3$ $\mu$Hz (black). The red curve shows the best fit to
the individual pulsation modes with Lorentzian profiles.}
\label{ThetaCyg_Q68_3panels}
\end{figure*}

The Fourier Transform of the Q6 data revealed a rich spectrum of overtones of solar-like oscillations, with excess power above the background in the frequency range $\sim$1200 to 2500\,$\mu$Hz.  Figure~\ref{ThetaCyg_Q68_3panels} shows the power-density spectrum of the processed Q6 and Q8 data.  The data show a large frequency separation $\Delta$$\nu$ of $\sim$84 $\mu$Hz with maximum oscillation amplitude at $\nu_{\rm max}$ = 1830 $\mu$Hz.  The appearance of the oscillation spectrum is very similar to that of other well-studied F stars such as Procyon~A \citep{Bedding10b,2015ApJ...813..106B}, HD49933 \citep{Appourchaux08, 2009A&A...507L..13B, 2012A&A...539A..63R}, HD181420 \citep{Barban09}, and HD181906 \citep{Garcia09}; see also Table 1 of \citet{2013A&A...550A.126M} and references therein.  The envelope of oscillation power is very wide, and modes are evidently heavily damped, meaning the resonant peaks have large widths in the frequency spectrum, which makes mode identification difficult \citep[see, e.g.,][and discussion in Section \ref{sec:ModeID}]{Bedding10a}.  For comparison to $\theta$ Cyg's $\nu_{\rm max}$ (1830 $\mu$Hz), the maximum in the Sun's power spectrum is at about 3150 $\mu$Hz, while for Procyon with mass 1.48 M$_{\odot}$ and luminosity 6.93 L$_{\odot}$, $\nu_{\rm max}$ is 1014 $\mu$Hz \citep{2011ApJ...743..143H}, and for HD49933, with mass 1.30 M$_{\odot}$ and luminosity 3.47 L$_{\odot}$, $\nu_{\rm max}$ is 1760 $\mu$Hz \citep{Appourchaux08}.



\section{New High-Resolution Spectra and Analyses}
\label{sec:spectroscopy}

A review of the extensive literature prior to the {\it Kepler} observations suggests that $\theta$ Cyg is a normal,
slowly rotating, solar-composition, F3V spectral-type star \citep{2003AJ....126.2048G} with $T_\mathrm{eff}$ around
6700$\pm$100~K and $\log g$ around 4.3$\pm$0.1 dex (see Appendix A).  This section summarizes analyses of high-resolution spectra taken subsequent to the {\it Kepler} observations by P.~I.~P\'{a}pics at the HERMES spectrograph\footnote{Supported by the Fund for Scientific Research of Flanders
(FWO), Belgium, the Research Council of KU Leuven, Belgium, the Fonds National
Recherches Scientific (FNRS), Belgium, the Royal Observatory of Belgium, the
Observatoire de Gen\`{e}ve, Switzerland and the Th\"{u}ringer Landessternwarte
Tautenburg, Germany} on the Mercator Telescope\footnote{Operated on the island of La Palma by the Flemish Community, at the Spanish Observatorio del
Roque de los Muchachos of the Instituto de Astrof\'{i}sica de Canarias} in May 2011, and by the team of D. Latham at the TRES spectrograph in December 2011.

\subsection{HERMES spectrum analyses of $\theta$ Cygni}

\begin{figure}
\center{\includegraphics[width=\columnwidth]{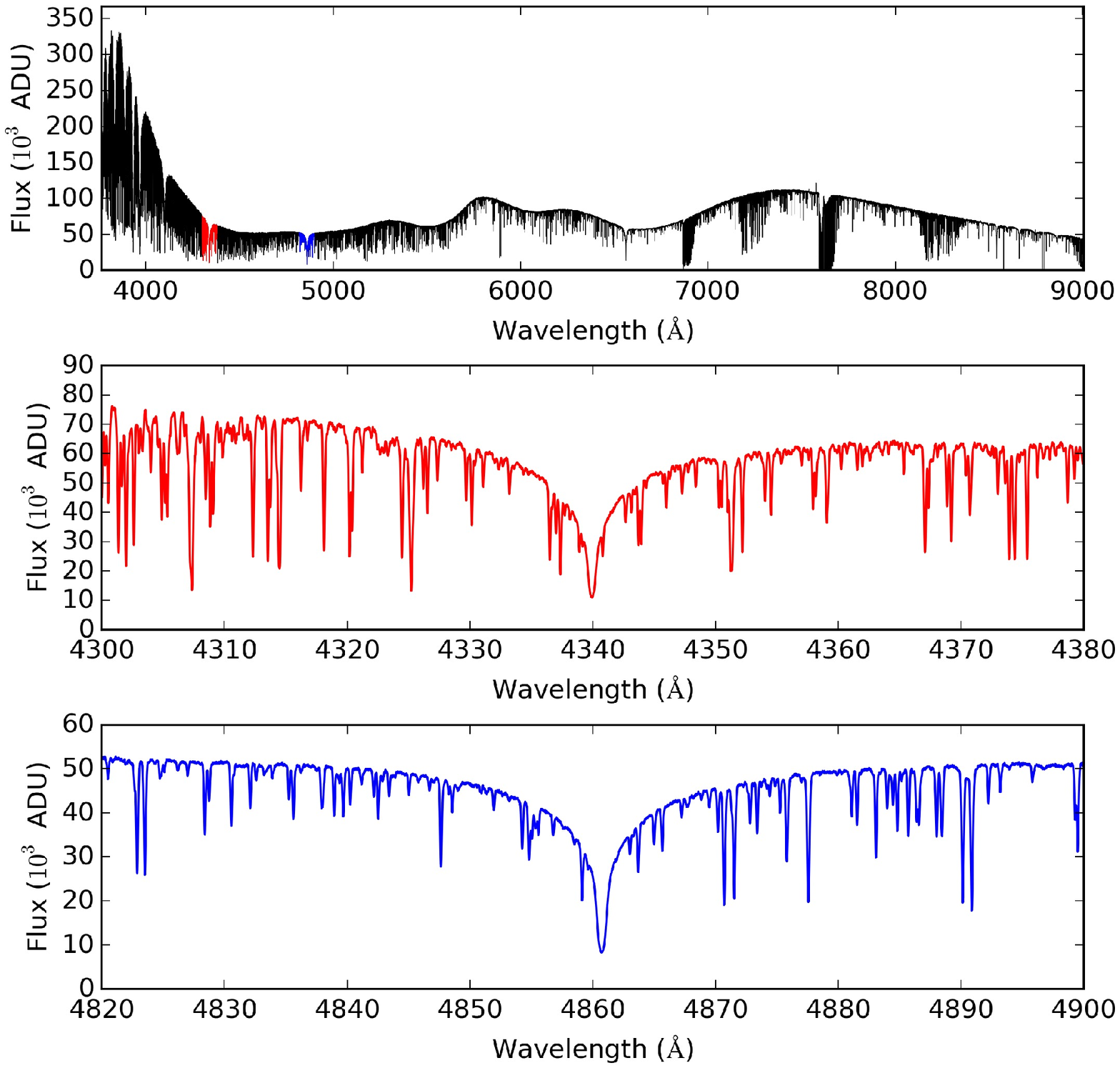}}
\caption{High-resolution spectrum of $\theta$ Cyg taken with HERMES Mercator spectrograph in May 2011 (top), and region around H$\gamma$ (middle) and H$\beta$ (bottom).} 
\label{HERMES}
\end{figure}

High-resolution high
signal-to-noise spectra were taken using the HERMES spectrograph
\citep{2011A&A...526A..69R} installed on the 1.2-meter Mercator telescope based
at the Roque de los Muchachos Observatory on La Palma, Canary Islands, Spain.
The spectrograph is bench-mounted and fiber-fed, and resides in a temperature-controlled enclosure to guarantee instrumental stability. During the
observations, HERMES was set to the HRF mode using the high-resolution fiber
with a spectral resolving power of $R=85,000$ delivering a spectral coverage from
377 to 900 nm in a single exposure and a peak efficiency of 28\%. The final
processed orders, along with merged spectra, were obtained on site using the
integrated HERMES data reduction pipeline.

Analyses of the HERMES spectrum (Fig.~\ref{HERMES}), discussed next, was undertaken independently by five of us, using
differing methods.

\subsubsection{{\sc vwa} }

The Versatile Wavelength Analysis ({\sc vwa}) method uses spectral synthesis to fit lines to determine their
equivalent widths (EWs). $T_{\rm eff}$ is found by adjusting it to remove any
slope in the Fe\,{\sc i} versus excitation potential of the lower level, using
lines with EW$<$100m{\AA}. The criterion for $\log{g}$ is that the average
abundance from the Fe\,{\sc i} and Fe\,{\sc ii} lines agree. In addition, checks
are made that Mg\,{\sc i}\,b and Ca lines at 6122{\AA} and 6162{\AA} are well
fitted. Since Van der Waals broadening is important for these lines, they have
been adjusted to agree with the solar spectrum for $\log{g}$ = 4.437
\citep{2010A&A...519A..51B}. The Fe abundance relative to solar, [Fe/H], is
calculated as the mean of Fe\,{\sc i} lines with EW$<$100~m{\AA} and
$>$5~m{\AA}. Microturbulence, ${v_{\rm mic}}$, is found by minimizing Fe\,{\sc i}
abundances versus EW, using only lines with EW$<$90~m{\AA}. Model atmospheres are an interpolation in
the {\sc MARCS} grid, with line lists from VALD \citep{1999A&AS..138..119K}.
Using a solar spectrum, each line has been forced to give the abundance in
\cite{2007SSRv..130..105G}, in order to give the correction to the $\log{gf}$ values.
Non-LTE effects are considered using \cite{1996A&A...312..966R}, since these effects
can be important for stars with $T_{\rm eff}$ above 6500~K.

\subsubsection{\sc uclsyn}

The analysis was performed based on the methods given in
\cite{2013MNRAS.428.3164D}. The {\sc uclsyn} code
\citep{1988eaa..conf...32S,Smith1992} was used to perform the analysis and
Kurucz {\sc atlas9} models with no overshooting were used
\citep{1997A&A...318..841C}. The line list was compiled using the VALD
database. The $H\alpha$ and $H\beta$ lines were used to give an initial
estimate of $T_{\rm eff}$. The $\log{g}$ was determined from the Ca~{\sc i} line
at 6439{\AA}, along with the Na~{\sc i} D lines.   Additional $T_{\rm eff}$ and $\log{g}$ diagnostics were performed using the Fe lines; however, the $T_{\rm eff}$ acquired from the excitation balance of the Fe~{\sc i} lines was found to be too high ($\sim$ 6900 K) and this $T_{\rm eff}$ was not used. A null dependence between the abundance and the equivalent width was used to constrain the microturbulence. The $\log{g}$ from the Fe lines was determined by requiring that the Fe~{\sc i} and Fe~{\sc ii} abundances agree, and the $T_{\rm eff}$ was also determined from the ionization balance.

The quoted error estimates include that given by the uncertainties in $T_{\rm
eff}$, $\log{g}$, and $v_{\rm mic}$, as well as the scatter due to measurement
and atomic data uncertainties. 

The projected stellar rotation velocity ($v \sin i$) was determined by fitting
the profiles of several unblended Fe\,{\sc i} lines in the wavelength range
6000--6200{\AA}. A value for macroturbulence of 6~km\,s$^{-1}$ was assumed,
based on slight extrapolations of the calibration by \citet{2010MNRAS.405.1907B}
and \citet{2014MNRAS.444.3592D}, and a
best-fitting value of $v \sin i = 4.0 \pm 0.4$~km\,s$^{-1}$ was obtained.

\subsubsection{\sc rotfit}

The {\sc rotfit} method is based on a $\chi^2$ minimization with a grid of
spectra of real stars with well-known astrophysical parameters
\citep{2006A&A...454..301F,2010ApJ...723.1583M,2013MNRAS.434.1422M}. Thus,
full spectral regions (discarding those ones heavily affected by telluric
lines), not individual lines, are used. The method derives $T_{\rm eff}$,
$\log{g}$, [Fe/H], $v \sin i$ and MK classification.

\subsubsection{{\sc SynthV}}

Stellar parameters ($T_{\rm eff}$, $\log{g}$, [M/H], $v_{\rm mic}$ and $v
\sin i$) are obtained by computing synthetic spectra and comparing them to the observed spectrum
\citep{2011A&A...526A.124L}. Atmosphere models were calculated with {\sc
LLmodels} \citep{2004A&A...428..993S}, the computation of synthetic spectra
was performed using {\sc SynthV} \citep{1996ASPC..108..198T}. Atomic data were
taken from VALD. The spectrum synthesis was done
on the wavelength range 4047--6849 {\AA}, covering both metal and the first four lines of the 
Balmer series. The local continuum of the observed  spectrum was corrected to
fit those of the synthetic  ones. $\chi^2$ statistics were used to determine
the optimum values of the atmospheric parameters and their errors based on
the 1-$\sigma$ confidence space in all parameters. 

\subsubsection{{\sc ares} {\rm +} {\sc moog}}

The stellar parameters were obtained from the automatic measurement of the
equivalent widths of Fe\,{\sc i} and Fe\,{\sc ii} lines with {\sc ares}
\citep{2007A&A...469..783S} and then imposing excitation and ionization
equilibrium using the {\sc moog} LTE line analysis code
\citep{1973PhDT.......180S} and a grid of Kurucz {\sc atlas}9 model atmospheres
\citep{1993KurCD..13.....K}. The Fe\,{\sc i} and Fe\,{\sc ii} line list
comprises more than 300 lines that were individually tested using
high-resolution spectra to check its stability to automatic measurement with
{\sc ares}. The atomic data were obtained from VALD, but with $\log{gf}$
adjusted through an inverse analysis of the Solar spectrum, in order to allow
for differential abundance analyses relative to the Sun
\citep{2008A&A...487..373S}. The errors on the stellar parameters are obtained
by quadratically adding 100~K, 0.13 and 0.06~dex to the internal errors on
$T_{\rm eff}$, $\log{g}$ and [Fe/H], respectively. These values were obtained by
considering the typical dispersion plotted in each comparison of parameters
presented in \cite{2008A&A...487..373S}. A more detailed discussion on the
errors derived for this spectroscopic method can be found in
\cite{2011A&A...526A..99S}.

\subsection{TRES Spectrum Analysis}

Two spectra were obtained with the Tillinghast Reflector \'Echelle
Spectrograph (TRES) on the 1.5-m Tillinghast Reflector at the Smithsonian's
Fred L. Whipple Observatory on Mount Hopkins, Arizona. The resolving power of
these spectra is 44,000, and the signal-to-noise ratio per resolution
element is 280 and 351 for one-minute exposures on BJD 2455905.568 and
2455906.544, respectively. The wavelength coverage extends from 385 to
909~nm, but only the three orders from 506 to 531 nm were used for the
analysis of the stellar parameters using Stellar Parameter Classification
\citep[SPC,][]{2012Natur.486..375B}, a tool for comparing an observed spectrum with
a library of synthetic spectra.  {\sc spc} is designed to solve simultaneously for
$T_{\rm eff}$, [M/H], $\log{g}$ and $v \sin{i}$. In essence,
{\sc spc} cross-correlates an observed spectrum with a library of synthetic
spectra for a grid of Kurucz model atmospheres and finds the stellar
parameters by determining the extreme of a multi-dimensional surface fit to
the peak correlation values from the grid.

The consistency between the {\sc spc} results for the two observations was
excellent, but undoubtedly the systematic errors are much larger, such as
the systematic errors due to the library of synthetic spectra. Based on past
experience, we assign floor errors of 50 K, 0.1 dex, 0.08 dex and
0.5 km\,s$^{-1}$ for $T_{\rm eff}$, $\log{g}$, [M/H] and $v \sin{i}$,
respectively. The library spectra were calculated assuming  $v_{\rm mic}$ =
2~km\,s$^{-1}$.  In tests of {\sc spc} it has been noticed that the $\log{g}$
values can disagree systematically with cases that have independent
dynamical determinations of the gravity for effective temperatures near 6500
K and above (e.g., for Procyon and Sirius). Therefore, as discussed in the next section, we have also used {\sc spc}
to determine $T_{\rm eff}$ and [M/H] whilst fixing $\log{g}$ to the value obtained from asteroseismology.

\begin{figure}
\center{\includegraphics[width=\columnwidth]{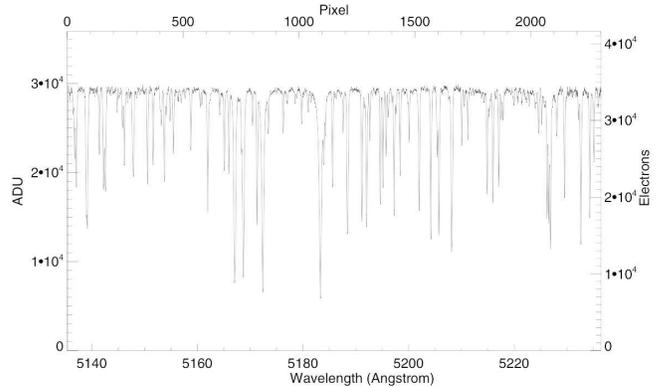}}
\caption{A spectrum of $\theta$ Cyg in the order containing the Mg b triplet, obtained with the TRES spectrograph on the 1.5-m reflector at the Fred Lawrence Whipple Observatory on Mount Hopkins, Arizona.  The resolving power is 44,000 and the SNR is 350 per resolution element of 6.8 km/s, at the center of the order.  The exposure time was 60 seconds.  The \'echelle blaze function has been removed by dividing with an exposure of a quartz iodine tungsten filament lamp.} 
\label{TRES}
\end{figure}

\begin{table*}
\begin{center}
\caption{Summary of the results from the spectral analyses}
\begin{tabular}{ccccccc}\hline
              & {\sc vwa}           & {\sc uclsyn}           & {\sc rotfit}            & {\sc SynthV}         & {\sc ares} + {\sc moog}            & {\sc spc} \\\hline
$T_{\rm eff}$ (K) &   6650 $\pm$ 80   &   6800 $\pm$ 108  &   6500 $\pm$ 150  &   6720 $\pm$  70  &  6942 $\pm$ 106  & 6637 $\pm$ 50 \\
$\log{g}$     &   4.22 $\pm$ 0.08 &   4.35 $\pm$ 0.08 &   4.00 $\pm$ 0.15 &   4.28 $\pm$ 0.19 &  4.58 $\pm$ 0.14 & 4.11 $\pm$ 0.1 \\
{[Fe/H]}      &$-$0.07 $\pm$ 0.07 &  +0.02 $\pm$ 0.08 &$-$0.2  $\pm$ 0.1  &$-$0.22 $\pm$ 0.05 &  0.08 $\pm$ 0.06 & $-$0.07 $\pm$ 0.08 \\
$v_{\rm mic}$ (km\,s$^{-1}$) &   1.66 $\pm$ 0.06 &   1.48 $\pm$ 0.08 &      n/a          &   1.93 $\pm$ 0.25 &  1.94 $\pm$ 0.10 &  (2.0) $\dag$\\
$v \sin{i}$ (km\,s$^{-1}$)  &      n/a          &   4.0  $\pm$ 0.4  &   4.0 $\pm$ 1.5   &   6.36 $\pm$ 0.61 &  n/a             & 7.0 $\pm$ 0.5 \\
\\
\multicolumn{7}{c}{fixing $\log{g} = 4.23\pm0.03$} \\
\\
$T_{\rm eff}$ (K)  &   6650 $\pm$ 80    &   6715 $\pm$ 92   &        n/a        &   6716 $\pm$  67  & 6866 $\pm$ 125   & 6705 $\pm$ 50 \\
{[Fe/H]}      &   $-$0.07 $\pm$ 0.07  &$-$0.03 $\pm$ 0.09 &        n/a        &$-$0.21 $\pm$ 0.05 & 0.06 $\pm$ 0.06  & $-$0.03 $\pm$ 0.08 \\
$v_{\rm mic}$ (km\,s$^{-1}$) &  1.66 $\pm$ 0.06          &   1.48 $\pm$ 0.08 &        n/a        &   1.92 $\pm$ 0.24 & 1.89 $\pm$ 0.10  &  (2.0) $\dag$\\
\hline
\end{tabular}
\label{summary}
\tablecomments{$\dag$ indicates an assumed value}
\end{center}
\end{table*}

\subsection{Stellar Parameters}
\label{SpectralResults}

\begin{figure*}
\includegraphics[width=\textwidth]{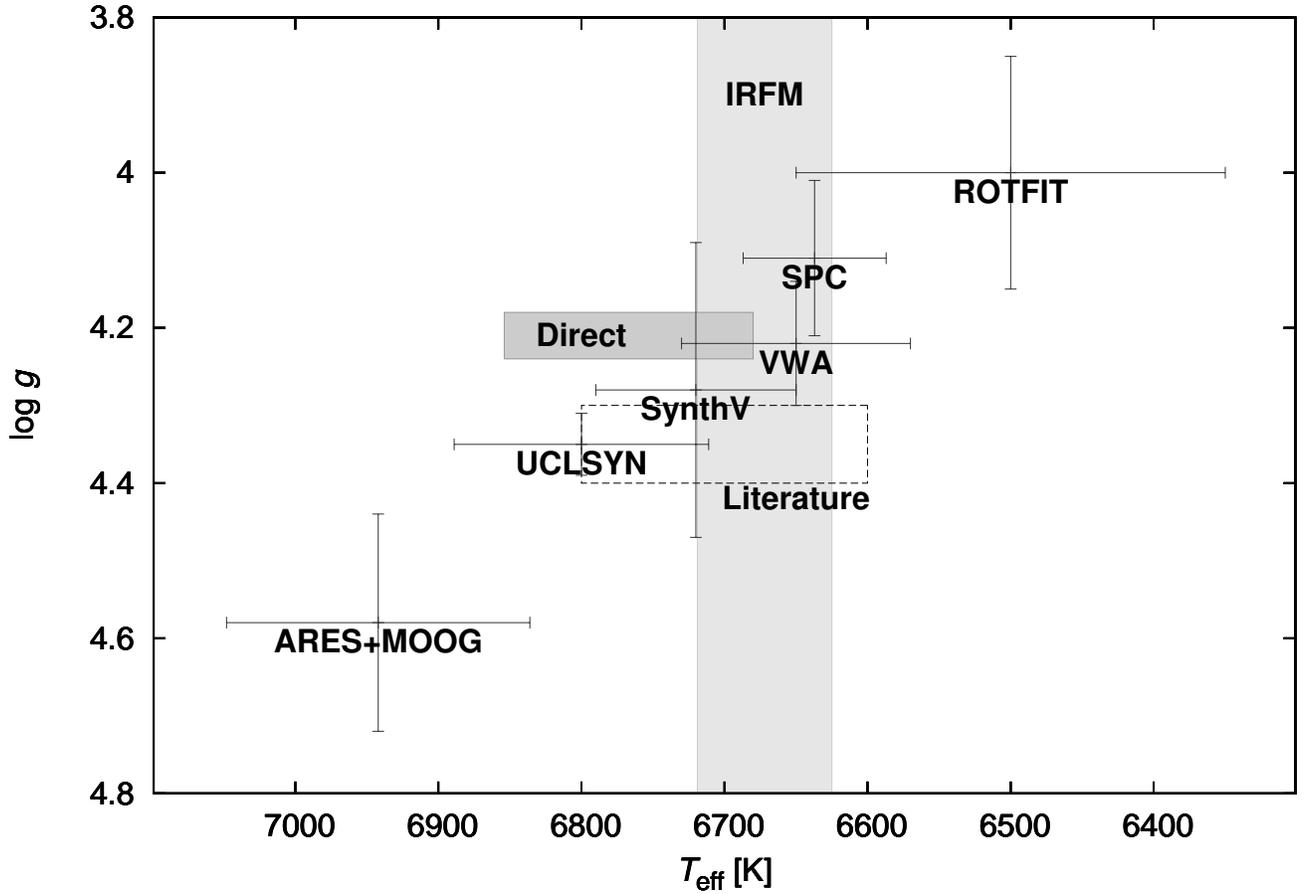}
\caption{Comparisons of results from the analyses of HERMES and TRES spectra.
The range of values of $T_{\rm eff}$ obtained by \cite{Blackwell98} using
the IRFM is shown by the light grey band. The interferometric $T_{\rm eff}$
from \cite{2012A&A...545A...5L} and the asteroseismic $\log{g}$ value are given as the dark grey box labelled `Direct'.
The dashed box indicates the most-probable parameters from the literature review.}
\label{results}
\end{figure*}

A summary of the results from the spectral analyses is given in
Table~\ref{summary}. There is a relatively large spread in the values of $T_{\rm
eff}$ and $\log{g}$ obtained from the spectral analyses. Examination of their
locations in the $T_{\rm eff}$--$\log{g}$ diagram (Fig.~\ref{results}), shows an apparent correlation between these two parameters.
This coupling between the two parameters is a known and common problem with spectral analyses, with some methods
more susceptible than others.

To address this degeneracy, the spectral analyses were, therefore, repeated using a fixed $\log{g} =
4.23 \pm 0.03$ derived from the interferometric and asteroseismic constraints on $\theta$ Cyg's mass and radius from \citet{2013MNRAS.tmp.1445W}, and which is also in line with the log $g$ values of the best-fit asteroseismic models discussed in Section \ref{sec:models}.  The exception is {\sc rotfit}, which due to its design for
use with a grid of real stars, cannot be used to derive parameters for a fixed
$\log{g}$. The results from the other methods are presented in the lower part of
Table~\ref{summary}. With the exception of {\sc ares + moog}, the model-atmosphere
spectroscopic methods all agree to within the error bars and differ by less than
70~K. The {\sc ares + moog} method is differential to the Sun, with a line list specifically
prepared for precise analysis of stars with temperatures closer to solar, and, therefore, $\theta$ Cyg
is too hot for this differential analysis.

It is interesting to explore why the model-independent {\sc rotfit} method is
giving slightly lower values. Fitting a spectrum to a grid of empirical spectra
of stars with known properties ought to give reliable results. The surface
gravity is higher than what would appear reasonable from external sources,
including the measured stellar luminosity. Inspection of figure~4 in
\cite{2011MNRAS.412.1210M} shows a similar difference at high $T_{\rm eff}$:
cooler $T_{\rm eff}$ and lower log $g$ compared to model atmosphere results by
$\sim$200~K and $\sim$0.2~dex, respectively. In fact, applying those corrections
would bring the {\sc rotfit} results into better agreement with the other spectroscopic
results.

From the remaining four spectral analyses, we obtain an average (after fixing log $g$) of $T_{\rm eff} =
6697 \pm 78$~K, where the error has been determined from quadrature sum of the
standard deviation of the average (31 K) and average of the individual methods'
errors (72 K). The latter is taken as a measure of the systematic uncertainty in
the temperature determinations \citep{2013ApJ...768...79G}. The result is consistent
with $T_{\rm eff} = 6672 \pm 47$~K from the IRFM
\citep{Blackwell98}, and with $T_{\rm eff} = 6767 \pm 87$~K 
 \citep{2012A&A...545A...5L} or $T_{\rm eff} = 6749 \pm 44$~K \citep{2013MNRAS.tmp.1445W} from interferometry.

\subsubsection{Metallicity}

The values for metallicity obtained from the spectroscopic analyses exhibit a
scatter of nearly 0.3~dex. In order to compare these we need to ensure that they
are all obtained relative to the same adopted solar value. The {\sc vwa} and {\sc ares + moog} methods are differential with respect to the Sun and provide a direct
determination of [Fe/H]. The {\sc uclsyn} and {\sc spc } analyses adopt the
\cite{2009ARA&A..47..481A} solar value of $\log A$(Fe) = 7.50, while the {\sc SynthV} analysis uses $\log
A$(Fe) = 7.45 \citep{2007SSRv..130..105G}. Adopting the
\cite{2009ARA&A..47..481A} solar Fe abundance would decrease the {\sc SynthV} value
to [Fe/H] = $-$0.26. While this value is discrepant from the other analyses, it
does agree with that found by {\sc rotfit} using empirical spectra. The average
metallicity from all the spectroscopic analyses, with the fixed $\log{g}$, is
[Fe/H] = $-0.07 \pm 0.12$~dex. If the {\sc SynthV} analysis is omitted, then the value
becomes [Fe/H] = $-0.02 \pm 0.06$~dex. Thus we conclude that $\theta$ Cyg
has a metallicity close to solar.

\subsubsection{Rotational Velocity}

The projected stellar rotational velocity ($v \sin{i}$) was determined by four of the methods. The {\sc uclsyn} analysis assumed a macroturbulence of 6~km\,s$^{-1}$ based on slight extrapolations of the calibrations by \cite{2010MNRAS.405.1907B} and \cite{2014MNRAS.444.3592D}, while the {\sc SynthV} and {\sc spc} analyses set macroturbulence to zero. The {\sc rotfit} method which uses spectra of real stars implicitly includes macroturbulence and agrees with the result of {\sc uclsyn}.  Setting macroturbulence to zero in the {\sc uclsyn} analysis yields $6.4 \pm 0.2$~km\,s$^{-1}$, which is in agreement
with {\sc SynthV} and {\sc spc}.  However, setting macroturbulence to zero is not a good assumption for slowly rotating stars, and leads to a large overestimation of  $v \sin i$ \citep[see][and references therein]{Murphy2016}.  Using Fourier techniques, \citet{1984ApJ...281..719G} obtained $v \sin i$ = 3.4 $\pm$ 0.4 km\,s$^{-1}$ and a macroturbulent velocity of 6.9 $\pm$ 0.3 km\,s$^{-1}$.  Given that we have not determined macroturbulence in our spectral analyses, we adopt Gray's values.

\section{Interferometric Radius}
\label{sec:interferometry}

$\theta$\,Cyg has also been the object of optical interferometry observations.  
\citet{2008ApJS..176..276V} used the Palomar Testbed Interferometer to identify 350 stars, including $\theta$\,Cyg, that are suitably pointlike to be used as calibrators for optical long-baseline interferometric observations.  They then used spectral energy distribution (SED) fitting (not the interferometry measurements) based on 91 photometric observations of $\theta$\,Cyg to derive a bolometric flux at the stellar surface and a bolometric luminosity, and estimate its angular diameter to be $0.760\,\pm\,0.021$\,milliarcsecond (mas).  Combining this angular diameter estimate
with the distance of 18.33 $\pm$ 0.05 pc given by the revised Hipparcos parallax 54.54 $\pm$ 0.15 mas \citep{vanLeeuwen2007b}, the derived radius of $\theta$ Cyg is $1.50\,\pm\,0.04$\,R$_{\odot}$.  

\citet{2012A&A...545A...5L} use observations from the VEGA/CHARA array to derive a limb-darkened angular diameter of 0.760 $\pm$ 0.003 mas, and a radius of 1.503 $\pm$ 0.007\,R$_{\odot}$.  \citet{2013MNRAS.tmp.1445W} use data from the Precision Astronomical Visual Observations (PAVO) combiner and the Michigan Infrared Combiner (MIRC) at the CHARA array, to derive a limb-darkened angular diameter of 0.753 $\pm$ 0.009 mas, and a radius of 1.48 $\pm$ 0.02 \,R$_{\odot}$.  A radius of $1.49 \pm 0.03 $R$_{\odot}$ encompasses both the \citet{2012A&A...545A...5L} and \citet{2013MNRAS.tmp.1445W} values.

Interferometry has the potential to constrain the radius of 
$\theta$\,Cyg more accurately than spectroscopy and photometry alone. It is notable that all three results, the \citet{ 2008ApJS..176..276V} estimate, and those reported in the two later observational papers, agree within their error bars on the angular diameter of $\theta$ Cyg, and that the inferred radius is constrained to better than would be obtainable without the interferometric observations.    If one were to use only the literature log $L/L_{\odot}$ = 0.63 $\pm$ 0.03 \citep{2008ApJS..176..276V} ($L = 4.26 \pm 0.30 L_{\odot}$) and $T_{\rm eff}$ 6745 $\pm$150K\citep{Erspamer03} and their associated error estimates to calculate the stellar radius,  the derived radius would be $1.53\,\pm\,0.13$\,R$_{\odot}$.  


\section{Large Separations, $\nu_{\rm max}$, and Estimating Stellar Parameters}
\label{sec:stellarparameters}

Solar-like oscillations with high radial orders exhibit characteristic large 
frequency separations, $\Delta\nu$, between modes of the same degree $l$ and consecutive radial order.  They also show 
small separations, $\delta \nu_{02}$ or $\delta \nu_{13}$, between $l=2$ and $l=0$, or between $l = 1$ and $l = 3$ modes of consecutive radial order, respectively \citep[see, e.g.,][]{Aerts2010}.

\begin{figure}
\center{\includegraphics[width=\columnwidth]{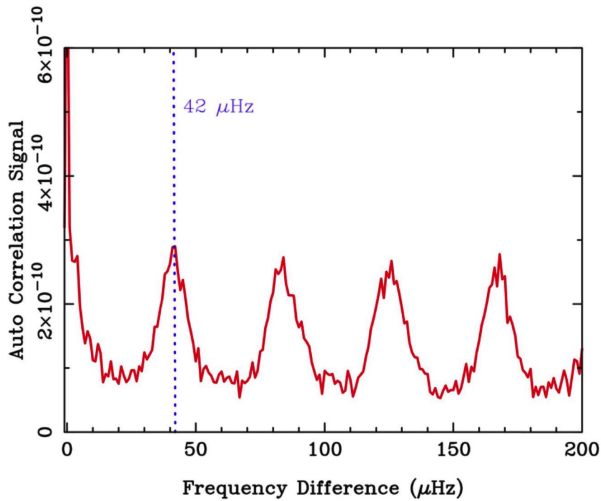}}
\caption{Autocorrelation of power spectrum, showing a 42-$\mu$Hz 
peak interpreted as half of the large frequency separation between modes.}
\label{autocorrelation} 
\end{figure}

An autocorrelation analysis of the frequency separations in the $\theta$\,Cyg 
solar-like oscillations first published by the {\it Kepler} team \citep{Haas2011} shows a peak at multiples of $\sim$42\,$\mu$Hz, interpreted to 
be half the large separation, $\frac{1}{2}\Delta\nu$ (Fig.~\ref{autocorrelation}). For comparison, 
half of the large frequency separation for the Sun is 67.5\,$\mu$Hz.

For solar-like oscillators, the frequency of maximum oscillation power, $\nu_{\rm max}$, has been found to scale as $gT_{\rm eff}^{-1/2}$ \citep{Brown1991, Kjeldsen1995, 2011A&A...530A.142B}, where $g$
is the surface gravity and $T_{\rm eff}$ is the effective temperature
of the star.  The most obvious spacings in the spectrum are the large
frequency separations, $\Delta\nu$.
These large separations scale to very good approximation as $\left< \rho \right>^{1/2}$,
$\left< \rho \right> \propto M/R^3$ being the mean density of the star
with mass $M$ and surface radius $R$  \citep[see, e.g.,][]{1986ApJ...306L..37U,Christensen-Dalsgaard1993}.

We used several independent analysis codes to obtain estimates of the
average large separation, $\left< \Delta\nu \right>$, and $\nu_{\rm
max}$, using automated analysis tools that have been developed, and
extensively tested \citep{Christensen-Dalsgaard2010, Hekker2010, Huber2009, Mosser2009, Mathur2010, Verner2011} for application to {\it Kepler} data \citep{Chaplin2011}. A final value of each parameter was selected by
taking the individual estimate that lay closest to the average over
all teams. The uncertainty on the final value was given by adding (in
quadrature) the uncertainty on the chosen estimate and the standard
deviation over all teams. The final values for $\left< \Delta\nu
\right>$ and $\nu_{\rm max}$ were $83.9 \pm 0.4\,\rm \mu Hz$ and $1829
\pm 54\,\rm \mu Hz$, respectively. 

We then provided a first estimate of the properties of the star using a
grid-based approach, in which properties were determined by searching
among a grid of stellar evolutionary models to get a best fit for the
input parameters, which were $\left< \Delta\nu \right>$, $\nu_{\rm
max}$, and the spectroscopically estimated $T_{\rm eff} = 6650 \pm
80\,\rm K$ and [Fe/H]=$-0.07 \pm 0.07$ of the star. Descriptions of
the grid-based pipelines used in the analysis may be found in \citet{Stello09, Basu2010, Gai2010, Quirion2011} and \citet{2014ApJS..210....1C}. The spread in the grid-pipeline results, which reflects
differences in, for example, the evolutionary models and input physics, was used to estimate the systematic uncertainties.

The oscillation power envelope of $\theta$\,Cyg (Fig.~\ref{ThetaCyg_Q68_3panels}) does not have
the typical Gaussian-like shape shown by cooler, Sun-like
analogues. Instead, it has a plateau, very reminiscent of the
extended, flat plateau shown by the oscillation power in the F-type
subgiant Procycon\,A, which has a similar $T_{\rm eff}$ \citep{Arentoft2008, Bedding10b}.  The shape of the envelope raises
potential questions over the robustness of the use of $\nu_{\rm max}$
as a diagnostic for the hottest solar-like oscillators.

Two sets of estimated stellar properties were returned by
each grid-pipeline analysis: one in which both $\Delta\nu$ and
$\nu_{\rm max}$ were included as seismic inputs; and one in which only
$\Delta\nu$ was used.

Both sets returned consistent results for the mass ($M=1.35 \pm
0.04\,\rm M_{\odot}$), $\log\,g$ ($4.208 \pm 0.006\,\rm dex$) and age
(${\rm \tau_{\odot}}=1.7 \pm 0.4\,\rm Gyr$), but not the
radius. There, using $\Delta\nu$ only yielded a radius of $R=1.51
\pm 0.02\,\rm R_{\odot}$, in good agreement with the interferometric value (Section \ref{sec:interferometry}),
while inclusion of $\nu_{\rm max}$ changed
the best-fitting radius to $R = 1.58 \pm 0.03\,\rm R_{\odot}$, an
increase of just under $2\sigma$ (combined uncertainty). This
difference -- albeit somewhat marginal -- could be reconciled by a
lower observed $\nu_{\rm max}$.

\section{Estimated mode linewidths and amplitudes}
\label{sec:damping}
\defcitealias{Gough77a}{Gough's (1977a,b)}
\defcitealias{Gough77b}{Gough's (1977a,b)}

\begin{figure}
\center{\includegraphics[width=0.99\columnwidth]{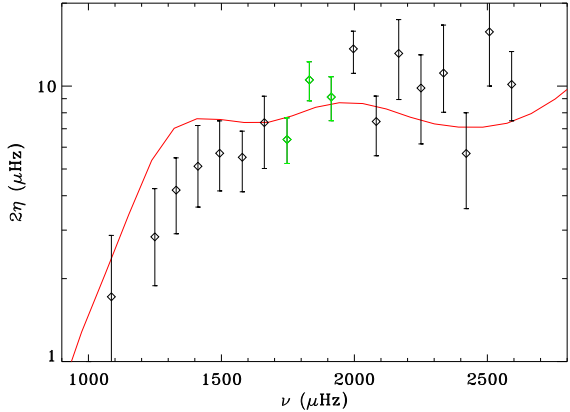}}
\caption{
Twice the theoretical linear damping rates for radial modes as a function of frequency
calculated for AMP Model 1 (Table \ref{AMPYRECModels}) with mass $M = 1.39$ M$_\odot$, luminosity
$L = 4.215$ L$_\odot$, effective temperature $T_{\rm eff} = 6753$ K, and
helium and heavy-element abundances by mass $X=0.7055$ and $Z=0.01845$ (solid red line).
The diamond symbols show the measured linewidths for observed radial modes of Table \ref{Frequencies}, with 3$\sigma$ error bars.  The green symbols indicate the three most prominent consecutive modes.}
\label{fig:damping-rates}
\end{figure}

We used theoretical calculations to estimate the linear damping rates $\eta(\nu)$ and amplitudes of the radial 
pulsation modes. The equilibrium and linear stability 
computations were similar to those by 
\citet[see also \citealt{Houdek99}]{Chaplin05}.
Convection was treated by means of a nonlocal, time-dependent generalization 
of \citetalias{Gough77a} mixing-length formulation.
The nonlocal formulation includes two more parameters, $a$ and $b$, in addition
to the mixing-length parameter, which control respectively the spatial 
coherence of the ensemble of eddies contributing to the turbulent fluxes 
of heat and momentum and the degree to which the turbulent fluxes are 
coupled to the local stratification. The momentum flux (turbulent pressure) 
was treated consistently in both the equilibrium and linear pulsation 
calculations. The mixing-length parameter was calibrated to obtain
the same surface convection-zone depth as suggested by the AMP evolutionary 
calculations
discussed in the Section \ref{subsection:AMP}. 
The nonlocal parameters, $a$ and $b$, were calibrated to reproduce 
the same maximum value of the turbulent pressure in the superadiabatic 
boundary layer as suggested by the grid results of three-dimensional (3D) 
convection simulations reported by \citet{2014MNRAS.445.4366T}.
\citetalias{Gough77b} time-dependent convection formulation includes also 
the anisotropy parameter $\Phi \equiv u_iu_i/w^2$, where $u_i=(u,v,w)$ is the 
convective velocity vector, for describing the anisotropy of the turbulent 
velocity field. In our model computations we varied $\Phi$ with stellar 
depth (Houdek et al. in preparation), guided by the 3D simulations by 
\citet{2014MNRAS.445.4366T}, and calibrated the value such as to obtain 
a good agreement between modelled linear damping rates and measured 
linewidths (see Figure~\ref{fig:damping-rates}). 
The remaining model computations were as described in \citet{Chaplin05}.

Figure~\ref{fig:damping-rates} shows twice the value of the theoretical 
linear damping rates (roughly equal to the full width at half maximum of 
the spectral peaks in the acoustic power spectrum) as a function of frequency 
for a model with the global parameters of AMP Model 1 of Table \ref{AMPYRECModels}.
The theoretical values are in good agreement with the range
of measured mean linewidths, 8.4 $\pm$ 0.3 $\mu$Hz, 
of the three most prominent modes (see Section~7 below).
near $\nu_{\rm max}\simeq1800\,\mu$Hz.

Amplitudes were estimated according to the scaling relation reported by 
\citet{Kjeldsen1995}, but also with the more involved stochastic
excitation model of \citet[see also \citealt{Houdek06}]{Chaplin05}. 
In this model
the acoustic energy-supply rate was estimated from the fluctuating Reynolds 
stresses adopting a Gaussian frequency factor and a Kolmogorov spectrum for 
the spatial scales \citep[see e.g.,][]{Houdek06, Houdek10, Samadi07}.
Kjeldsen \& Bedding's scaling relation suggest a maximum luminosity (intensity) 
amplitude of about $2.2$ times solar, which is in reasonable agreement with the
observed value of $4-5\,$ppm, assuming a maximum solar amplitude of $2.5\,$ppm
\citep{Chaplin2011}.
The adopted stochastic excitation model provides a maximum amplitude of
about $2.6$ times solar, which is slightly larger than the value from the
scaling relation. The overestimation of pulsation amplitudes in relatively
`hot' stars has been reported before, for example, for Procyon A,
\citep[see, e.g.,][]{Houdek06, 2010A&ARv..18..197A}.  Note that the most recent amplitude-scaling relation anchored on open-cluster red giants \citep{2011ApJ...737L..10S}, which agrees with observations of main-sequence stars \citep{2011ApJ...743..143H}, predicts 5.1 ppm for $\theta$ Cyg, in good agreement with its observed value.

\section{Mode Identification and Peak Bagging}
\label{sec:ModeID}

\subsection{\'Echelle diagram}
\label{subsection:echellediagram}

A convenient way to visualise solar-like oscillations is with the \'echelle diagram \citep{Grec83},
which makes use of the nearly-regular pattern exhibited by the modes. In these diagrams, the power spectrum
is split up into slices of width $\Delta\nu$, which are stacked on top of each other. Modes of the same angular 
degree $l$ form nearly vertical ridges in these diagrams. The \'echelle diagram for $\theta$~Cyg is shown in 
Figure \ref{thCygechelle}. The width of the \'echelle diagram is the large separation, $\Delta\nu=83.9$  $\mu\mathrm{Hz}$.  The \'echelle diagram can be useful for finding weak modes that fall along the ridges, and also for making the mode identification, that is, determining the $l$ value of each mode.

In stars like the Sun, the mode identification can be trivially made from the \'echelle diagram because the $l=0$ 
and $l=2$ modes form a closely spaced pair of ridges that is well-separated from the $l=1$ modes. However, in 
hotter stars we see stronger mode damping, leading to shorter mode lifetimes and larger linewidths 
\citep{Chaplin09, Baudin11, Appourchaux12a, 2013MNRAS.430.2313C}. This blurs the $l=0,2$ pairs into a single ridge that is very similar in appearance to the $l=1$ ridge. This problem was first observed in the CoRoT F star HD\,49933 
\citep{Appourchaux08} and subsequently in other CoRoT stars \citep{Barban09, Garcia09}, Procyon \citep{Bedding10b}, 
and many {\it Kepler} stars \citep[e.g.,][]{2012ApJ...749..152M, 2012A&A...543A..54A,2014ApJS..214...27M}. From Figure \ref{thCygechelle} it is clear that $\theta$ Cyg also suffers from this problem. Without a clear mode identification, the prospects for asteroseismology on this target are severely impeded.

Several methods have been proposed to resolve this mode identification ambiguity from \'echelle diagrams. One method is to attempt to fit both possible scenarios. A more likely fit should arise for the correct identification as it will better account for the 
additional power provided by the $l=2$ modes to one of the ridges. However, this method can run into difficulties
with low signal-to-noise observations, or with short observations in which the Lorentzian mode profiles have not
been well resolved. Rotational splitting, as well as wide linewidths and short lifetimes, will also create complications for this method.

Despite these difficulties, we attempted to fit the two possible mode identifications (or scenarios) using the Markov Chain Monte Carlo (MCMC) method and a Bayesian framework. A MCMC algorithm performs a random walk in the parameter space and explores the topology of the posteriori distribution (within bounds defined by the priors).  This method enabled us to determine the full probability distribution of each of the parameters and to determine the so-called evidence \citep[e.g.][]{2009A&A...506...15B, 2009A&A...507L..13B}. The evidence for the two mode identifications can be compared in order to evaluate the odds of the competing scenarios (hereafter referred as scenario A and scenario B) in terms of probability. Scenario A corresponds to $l=0$ at $\approx 1038$ $\mu$Hz (or $\varepsilon$ is 1.4), and scenario B corresponds to $l=0$ at $\approx 1086$ $\mu$Hz (or $\varepsilon$ is 0.9). With a probability of 70\%, we found that scenario B is only marginally more likely.

\subsection{$\varepsilon$ parameter}
\label{subsection:epsilon}
\begin{figure*}
\mbox{
\includegraphics[scale=0.52]{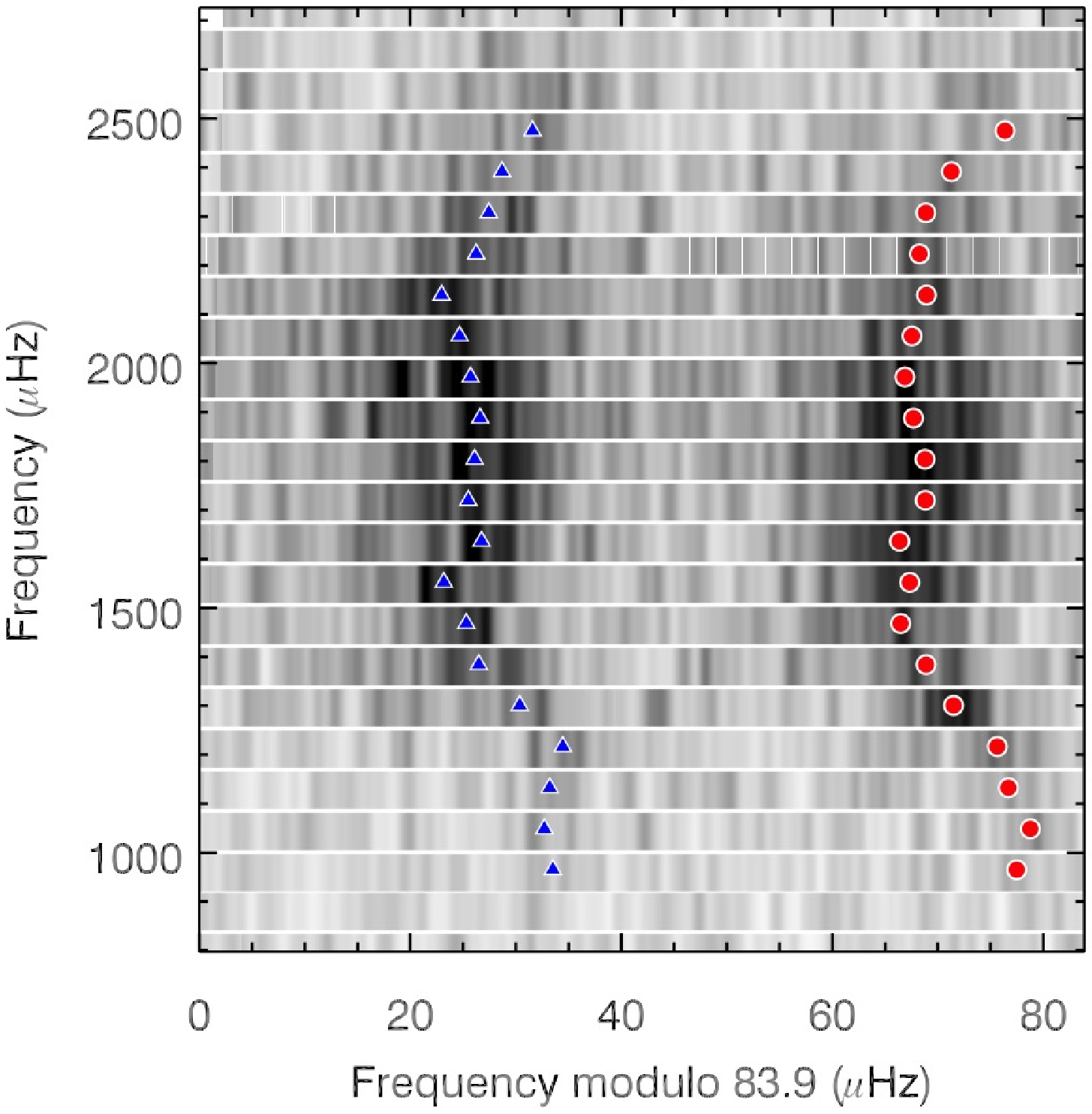}
\includegraphics[scale=0.52]{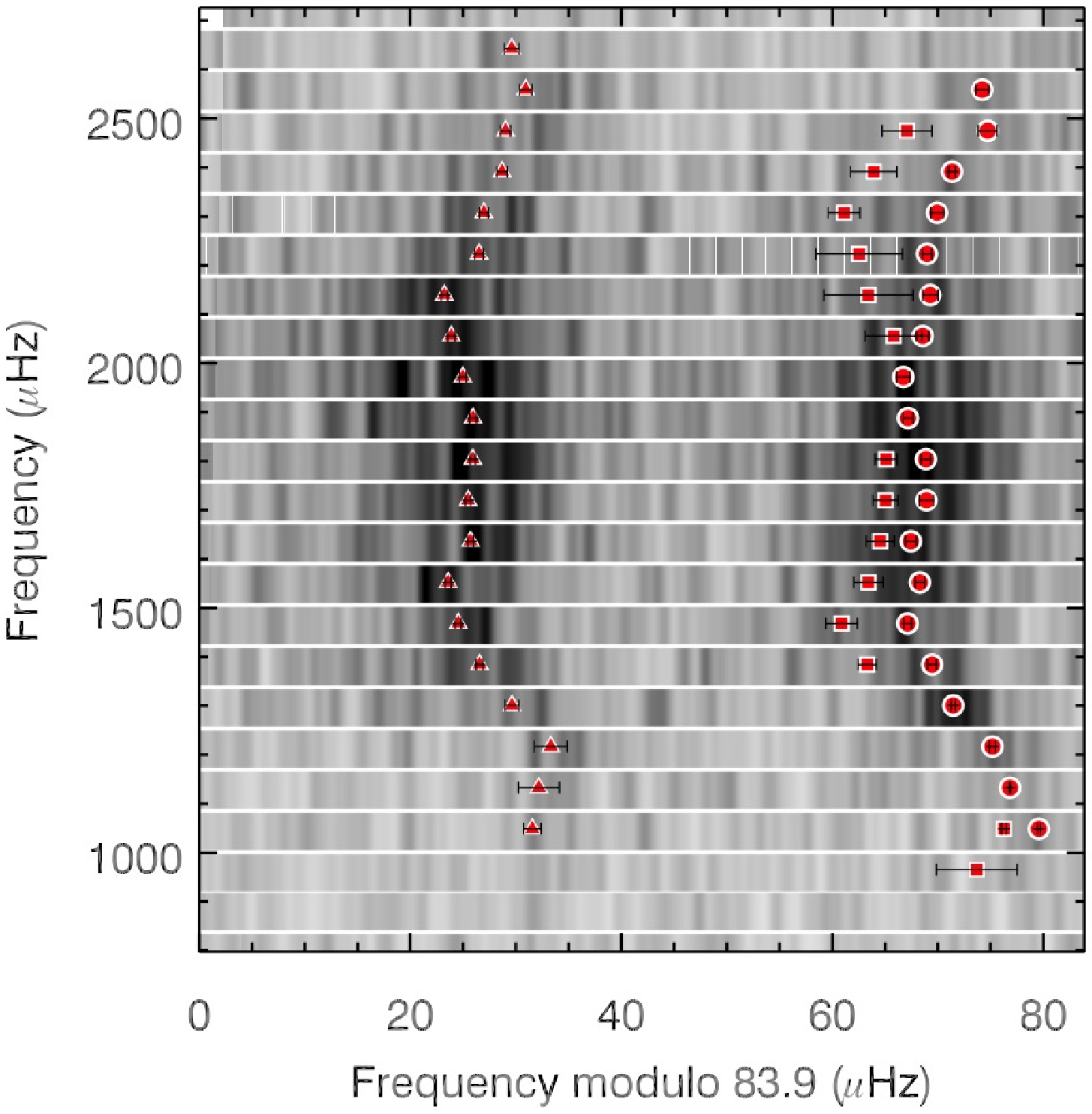}
}
\caption{\'Echelle diagram of $\theta$ Cyg. Left:  The blue triangles and red circles show the central frequencies along each ridge in the diagram. Blue triangles correspond to the $l$ = 0 ridge in Scenario~A, while red circles correspond to the $l$ = 0 ridge in Scenario~B. Right:  \'Echelle diagram of $\theta$ Cyg showing identified frequencies for Scenario B in red.  Modes are identified as $l$= 0 (circles), $l$= 1 (triangles), and $l$= 2 (squares). For reference in both figures, a smoothed gray-scale map of the power spectrum is shown in the background.}
\label{thCygechelle}
\end{figure*}

An alternative method has been introduced by \citet{2012ApJ...751L..36W} following on from work by \citet{Bedding10b}, 
which uses the absolute mode frequencies, as encoded in the parameter $\varepsilon$. The value of $\varepsilon$ is determined 
by the phase shifts of the oscillations as they are reflected at their upper and lower turning points. In the \'echelle 
diagram, $\varepsilon$ can be visualized as the fractional position of the $l=0$ ridge across the diagram. The left ridge 
in Figure \ref{thCygechelle} is approximately 40\% across the \'echelle diagram, so if this ridge is due to $l=0$ modes then 
the value of $\varepsilon$ is 1.4. We will refer to this as Scenario A. Alternatively, if the right ridge is due 
to $l=0$ modes (Scenario B), then the value of $\varepsilon$ is 0.9, since this ridge is approximately 90\% across the
\'echelle diagram. If it is known which value $\varepsilon$ should take, then the correct mode identification will be known.

It has been found that a relationship exists between $\varepsilon$ and effective temperature, $T_\mathrm{eff}$, both in 
models \citep{White11a} and observationally \citep{White11b}. Furthermore, since a relation also exists between
$T_\mathrm{eff}$ and mode line width, $\Gamma$ \citep{Chaplin09, Baudin11, Appourchaux12a, 2013MNRAS.430.2313C}, there is also a
relation between $\varepsilon$ and $\Gamma$ \citep{White11b}. Given these observed relationships between $\varepsilon$, 
$T_\mathrm{eff}$ and $\Gamma$ measured from an ensemble of stars, and the measured values of $T_\mathrm{eff}$ and 
$\Gamma$ in $\theta$~Cyg, the likelihood of obtaining either possible value of $\varepsilon$ ($\varepsilon_\mathrm{A}$ 
and $\varepsilon_\mathrm{B}$) can be calculated.

\begin{figure*}
\mbox{
\includegraphics[scale=0.52]{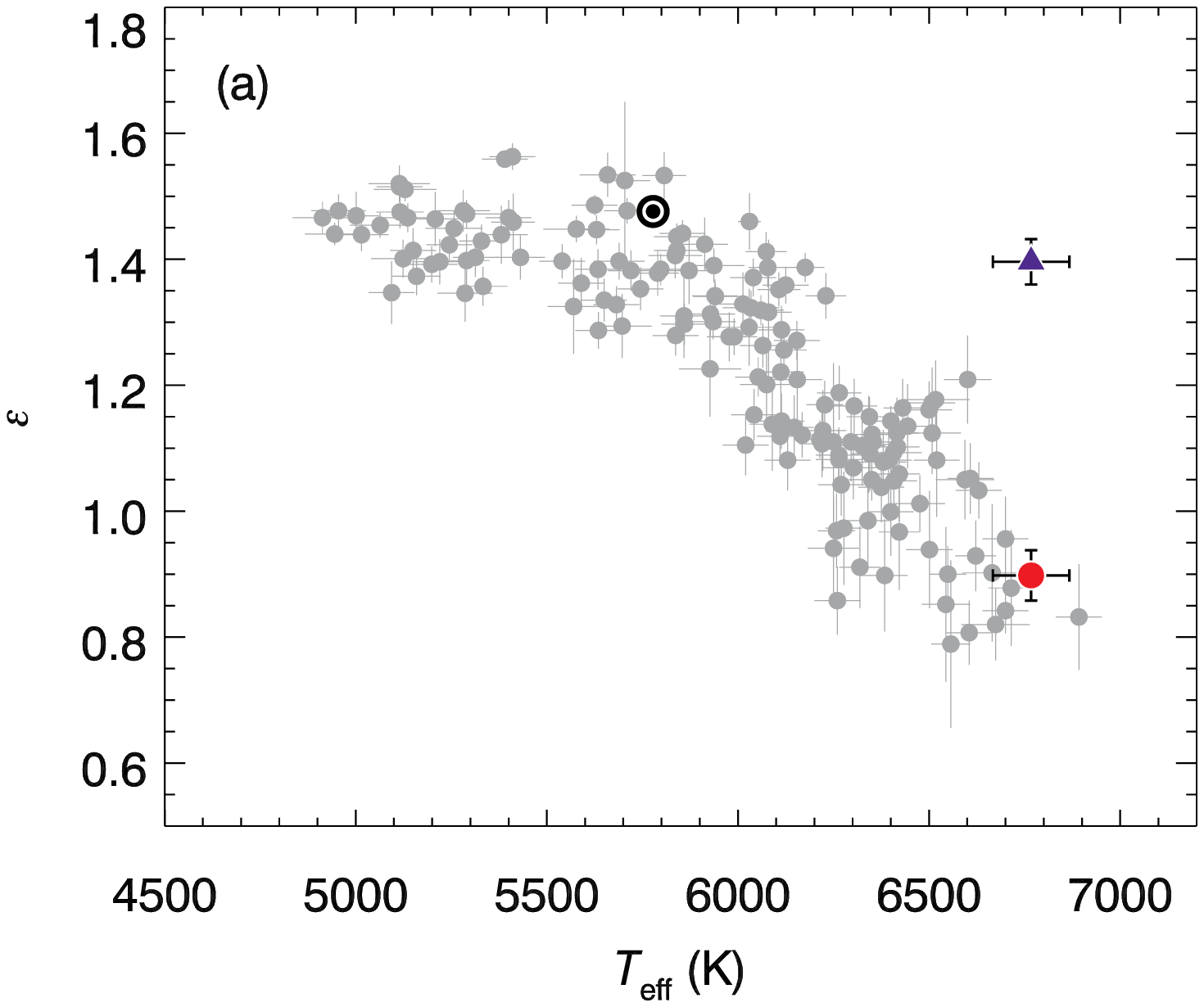}
\includegraphics[scale=0.52]{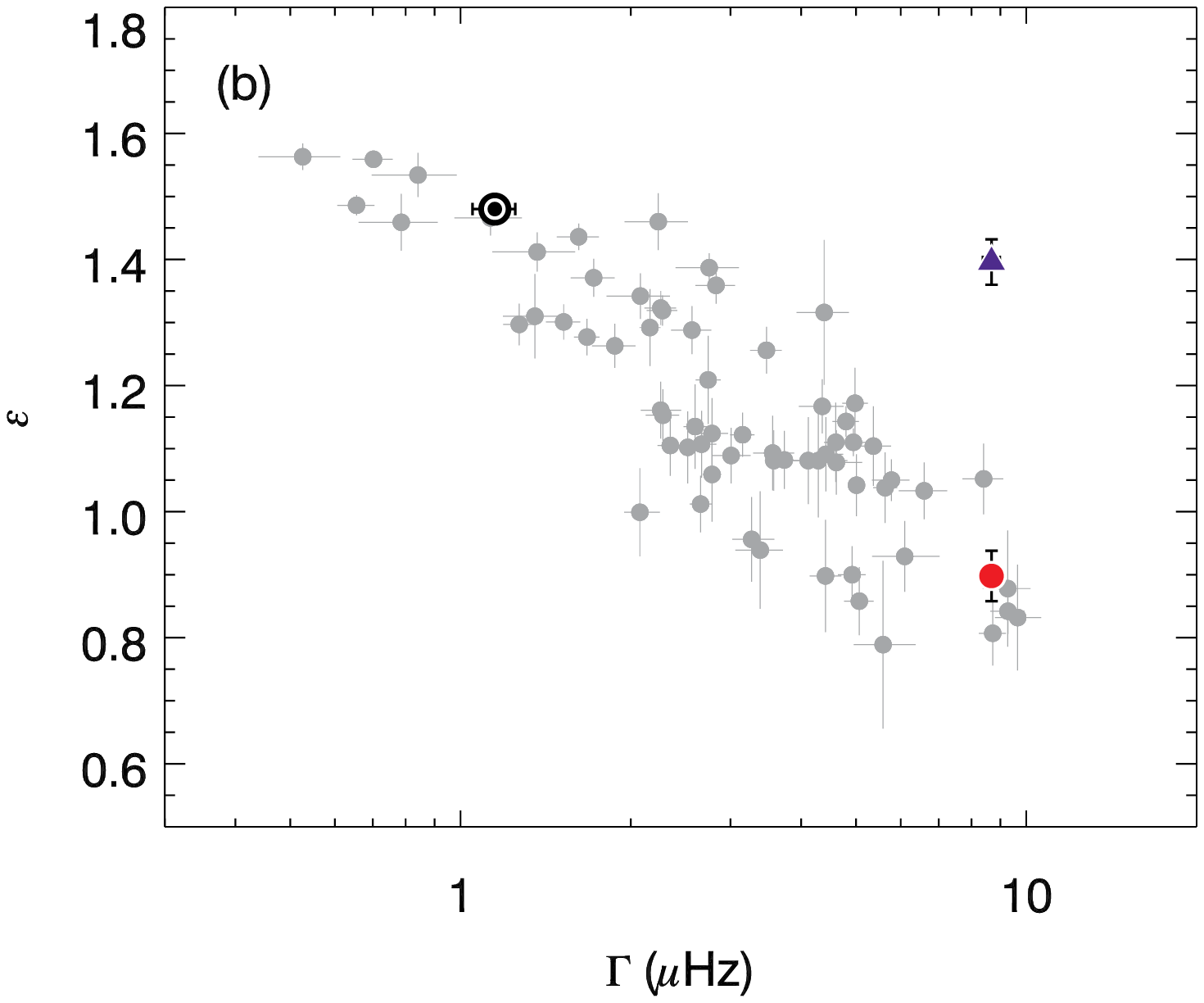}
}
\caption{The possible locations of $\theta$~Cyg (blue triangles for Scenario A and red circles for Scenario B) in the (a) $\varepsilon$ -- $T_\mathrm{eff}$ plane and (b) $\varepsilon$ -- $\Gamma$ plane.  Grey points are {\it Kepler} stars from \citet{2012ApJ...751L..36W}. 
The Sun is marked by its usual symbol.  This figure shows that Scenario B is the more likely of the two discussed in Section \ref{sec:ModeID}, since an $\varepsilon$ value of $\sim$ 0.9 (as opposed to $\sim$ 1.4) for $\theta$ Cyg places it in line with the other {\it Kepler} stars.}
\label{epsilon}
\end{figure*}

Following the method of \citet{2012ApJ...751L..36W}, we measured the ridge
frequency centroids from the peaks of the heavily smoothed power spectrum. We perform a linear least-squares fit to
the frequencies, weighted by a Gaussian window centered at $\nu_{\rm max}$ with FWHM of 0.25\,$\nu_{\rm max}$,
to determine the values of $\Delta\nu$, $\varepsilon_\mathrm{A}$ (1.40$\pm$0.04) and $\varepsilon_\mathrm{B}$ (0.90$\pm$0.04). 
The average linewidth, $\Gamma$ of the three highest amplitude modes is 8.4 $\pm$ 0.3 $\mu$Hz. The positions of 
$\theta$ Cyg in the $\varepsilon$ -- $T_\mathrm{eff}$ and $\varepsilon$ -- $\Gamma$ planes are shown in Fig.~\ref{epsilon}.
We find the most likely scenario to be Scenario B, with a probability of $99.9$\% (calculated in a Bayesian framework described in \cite{2012ApJ...751L..36W}).  According to \citet{2013A&A...550A.126M}, who compare the asymptotic and global seismic parameters, only scenario B with $\varepsilon \simeq 0.9$ is possible for a main-sequence star as massive as $\theta$ Cyg.  Table~\ref{Frequencies} lists the frequencies for that most likely scenario.

\subsection{Peak bagging}
\label{subsection:peak bagging}

Individual pulsation frequencies probe the stellar interior, so that by taking them into account, it is possible to improve the precision on the global fundamental parameters of the star.
This however requires to measure these frequencies precisely and accurately using the so-called peak-bagging technique. Peak bagging could be performed using several statistical methods.
The most common is the Maximum Likelihood Estimator (MLE) approach \citep{1990ApJ...364..699A} and has been thoroughly used to analyze the low-degree global acoustic oscillations of the Sun \citep[see, e.g.,][]{1996MNRAS.280..849C}. 

Although fast, the MLE is only suited in cases where the likelihood function has a well defined single maximum so that convergence towards an unbiased measure of the fitted parameters is ensured \citep{1998A&AS..132..107A}.  Unfortunately, stellar pulsations often have much lower signal-to-noise ratio than solar pulsations, so that the likelihood may have several local maxima. In this situation, the MLE may not converge towards the true absolute maximum of probability.

Conversely, the Bayesian approaches that rely on sampling algorithms such as the MCMC do not suffer from convergence issues \citep{2009A&A...506...15B, 2009A&A...507L..13B, 2011A&A...527A..56H}. This is because whenever local maxima of probability exist, these are sampled and become evident on the posterior probability density function of the fitted parameters.

In order to get reliable estimates of the mode frequencies for Scenario B, we choose to use such a Bayesian approach. The power spectrum of each star was modelled as a sum of Lorentzian profiles, with frequency, height and width as free parameters. The fit also included the rotational splitting and the stellar inclination  as additional free parameters. The noise background function was described by the sum of two Harvey-like profiles \citep{1985ESASP.235..199H} plus a white noise.  Table~\ref{Frequencies} lists the median of the frequencies obtained from the fit the MCMC algorithm, along with the $1\sigma$ uncertainty.

\section{Stellar Models Derived from $p$ Modes and Observational Constraints}
\label{sec:models}

We explored seismic models for $\theta$ Cyg matching the constraints from spectroscopic and interferometric constraints, as well as the $p$-mode frequencies and mode identifications derived from the {\it Kepler} data, using several different methods and stellar evolution and pulsation codes, as described below.

\subsection{Results from YREC Stellar Modeling Grid}
\label{subsection:YREC}

We use the Yale Rotating Stellar Evolution Code, YREC \citep{Demarque2008}, to calculate a grid of stellar models and their frequencies using a Monte Carlo algorithm to survey the parameter space constrained by the $\theta$ Cyg spectroscopic and interferometric observations summarized in Table \ref{AMPYRECModels}.  This Yale Monte Carlo Method (YMCM) is described in more detail by \cite{2015MNRAS.452.2127S}.  The Scenario B frequencies of Table~\ref{Frequencies} are used as seismic constraints.   The models are constructed using the OPAL equation of state \citep{Rogers02}, OPAL high-temperature opacities \citep{Iglesias1996}, and \citet{2005ApJ...623..585F} low-temperature opacities. Nuclear reaction rates are from \citet{Adelberger08} except for the $^{14}$N$(p,\gamma)^{15}$O reaction for which the rate of \citet{Formicola2004} is adopted. Convection was treated using the mixing-length formalism of \citet{Bohm-Vitense58}.  Models are constructed with a core overshoot of 0.2 pressure scale heights ($H_p$) unless the convective core size is less than $0.2~H_p$, in which case no overshoot is used.  Oscillation frequencies are calculated using the code described by \cite{1994A&AS..107..421A}.

Modeling $\theta$ Cyg poses the usual challenges for an F star. The outer convection zone is relatively thin compared to that of the Sun, which means that unless diffusive settling
is switched off, or artificially slowed down, the model soon loses most or all of the helium and metals at the photosphere. As a result 
models were constructed assuming that the gravitational settling of helium and heavy elements 
is too slow to affect the models.

These YMCM models use the surface-term correction of \cite{2014A&A...568A.123B}.  The surface term is the frequency-dependent deviation of model frequencies from the observed ones and is caused predominantly because of our inability to model the surface of stars
properly. The main shortcoming of the models arises because the effect of turbulence is 
not included. In the solar case the surface term causes model frequencies to
be larger than observed frequencies \citep[see, e.g.,][]{Christensen-Dalsgaard1996}. The 
frequencies of the low-frequency modes match observations, while those of high-frequency
modes are larger than the observed ones. It is usually assumed that the
surface term for models of stars other than the Sun can be simply scaled from the solar case \citep{Kjeldsen2008,2014A&A...568A.123B}.  For future work, 3-D hydrodynamical modeling \citep{2015A&A...583A.112S} could be used to constrain the surface-effect corrections.

The best-fit models are identified by calculating a $\chi^2$ value for the seismic and spectroscopic quantities separately, and adding them together.  A likelihood is defined using the total $e^{(-\chi^2)}$ and then used as a weight to find the mean and standard deviation of the model properties.   Table\,~\ref{AMPYRECModels} summarizes the mean and standard deviations of properties of the models, as well as the properties of the best-fit (highest likelihood) model. Figure \ref{AMPYREC_Echelle} shows the \'echelle diagram for this best-fit model compared to the observed frequencies.

\subsection{Results from AMP Stellar Model Grid Optimization Search}
\label{subsection:AMP}

The Asteroseismic Modeling Portal \citep[AMP, ][]{2009ApJ...699..373M} searches for models that minimize the average of the $\chi^{2}$ values for both the seismic and spectroscopic constraints.  The AMP has been applied extensively to modeling of other {\it Kepler} targets \citep[e.g.,][]{2012ApJ...749..152M, 2012ApJ...748L..10M}.  Although we ran many models exploring various optimization schemes and the effects of diffusive settling, we present results only for models without diffusive settling of helium or heavier elements, as the models including helium settling produce an unrealistic surface helium abundance, and AMP models do not (yet) include diffusion of heavier elements.  As noted in Section \ref{subsection:YREC} above, the envelope convection zone in F stars is shallow enough that most of the helium and metals would diffuse from the surface when diffusion is included; since we observe a non-zero metallicity at the surface of $\theta$ Cyg, it follows that some mechanisms, such as convective mixing or radiative levitation, are counteracting diffusive settling.  However, it is not physically correct to turn off diffusive settling completely, as evidence from helioseismology supports diffusive settling in the Sun \citep[see, e.g.,][]{1993ApJ...403L..75C, 2005ApJ...627.1049G}.

The AMP search makes use of an option that optimizes the fit to the frequency separation ratios defined by \cite{2003A&A...411..215R}, as well as to the individual frequencies using the empirical surface correction of \cite{Kjeldsen2008}.  The fit to the frequencies is also weighted to de-emphasize the highest frequency modes that are most affected by inadequacies in modeling the stellar surface. For complete details, see \cite{2014ApJS..214...27M}.  The models use the OPAL \citep{Iglesias1996} opacities and \citet{gn93} abundance mixture, and do not include convective overshooting.

\begin{figure*}
\includegraphics[width=\columnwidth]{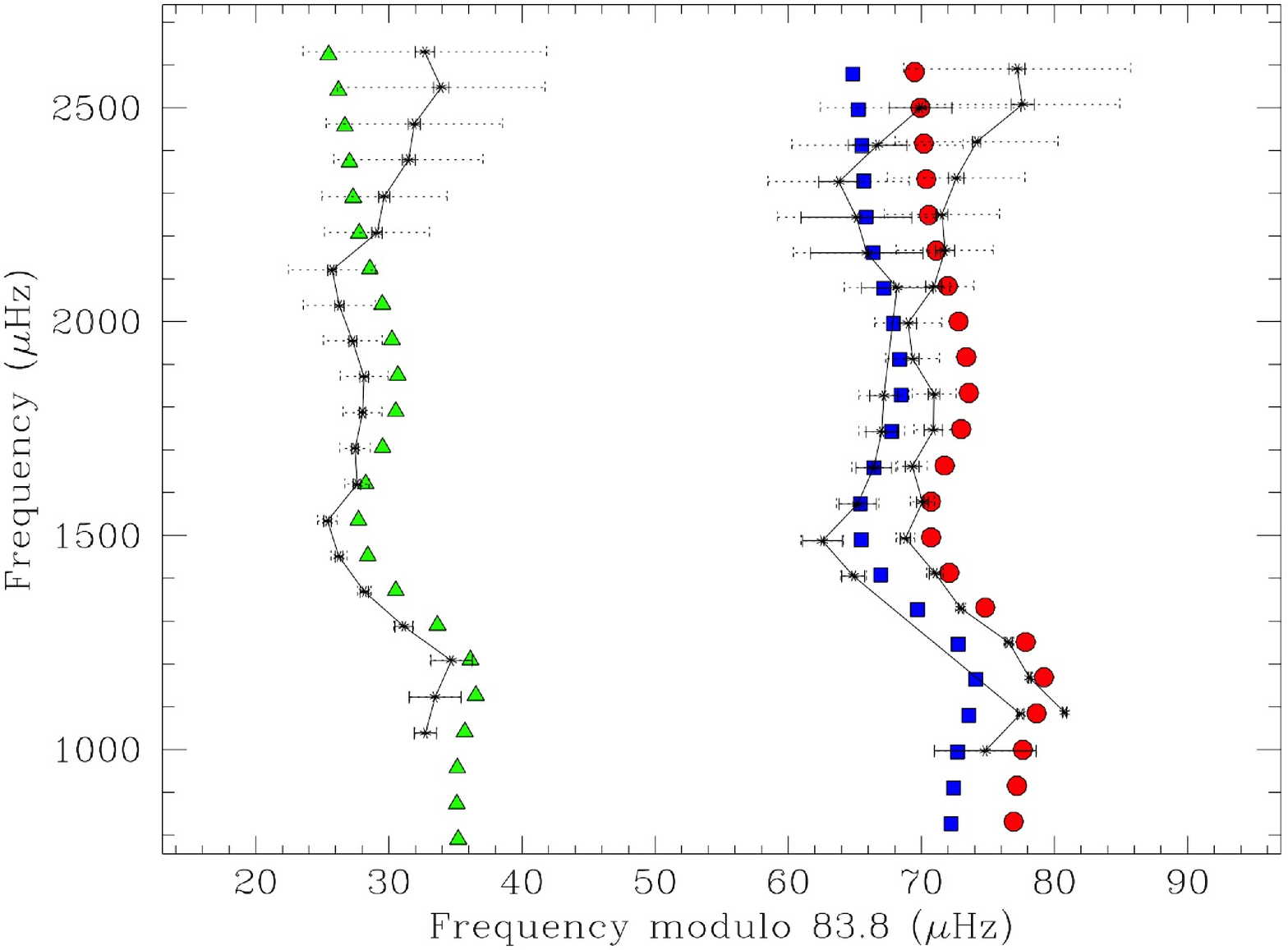}
\includegraphics[width=\columnwidth]{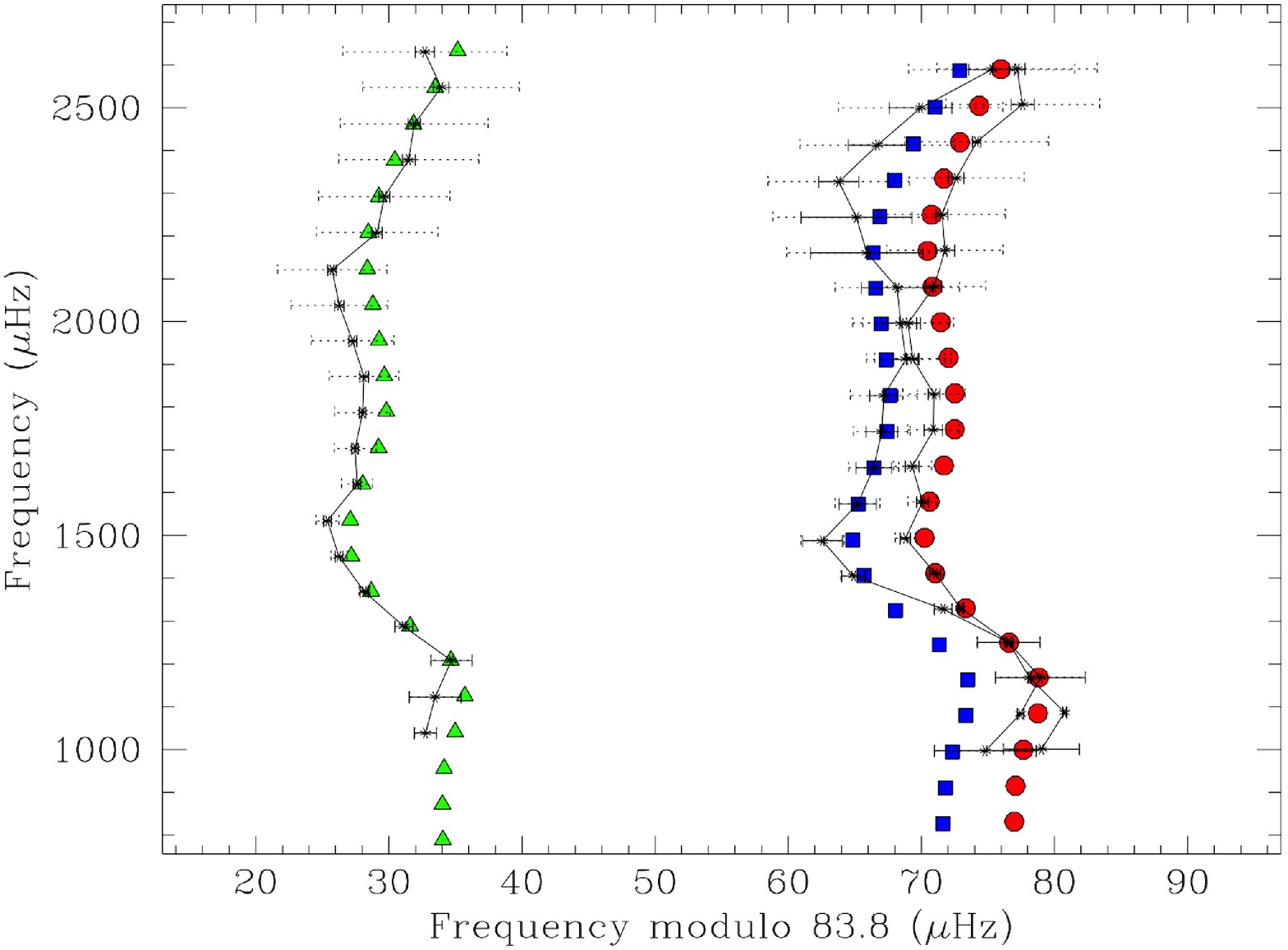}
\center{\includegraphics[width=\columnwidth]{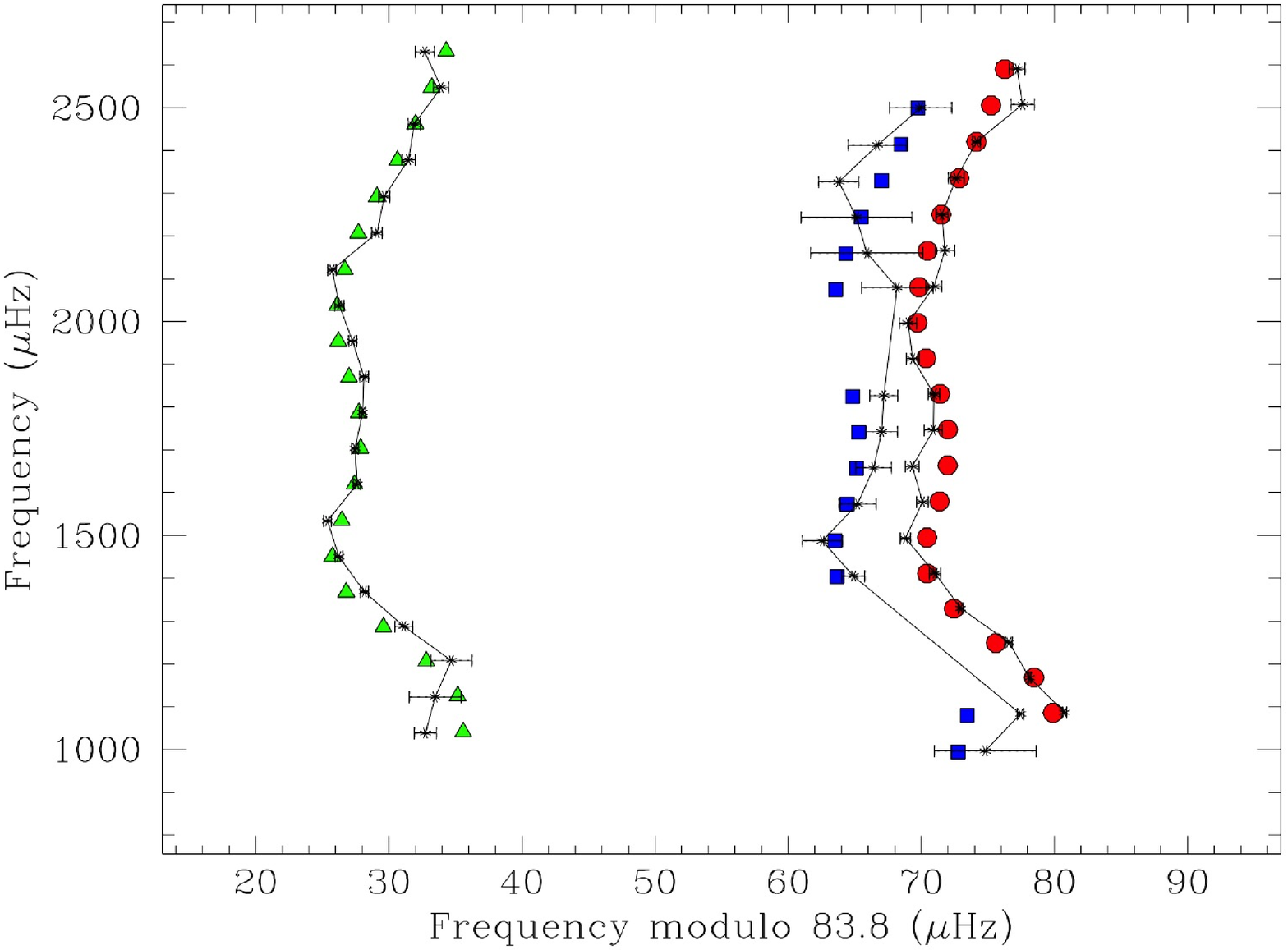}}
\caption{\'Echelle diagrams comparing the observed oscillation frequencies (connected
points) with calculated frequencies of AMP Model 1 and Model 2 (top, left and right, respectively) and best-fit YREC model (bottom).  The frequencies are derived from {\it Kepler} Q6 and Q8 data (Table \ref{Frequencies}).  Solid error bars indicate the observational uncertainties for each frequency.  Colored symbols show the radial ($\circ$), dipole ($\bigtriangleup$), and
quadrupole ($\Box$) modes after applying an empirical surface correction.  For the AMP models, dotted error bars show the
effective uncertainties adopted for the modeling, which treats the surface correction as a systematic error in the model.  AMP Model 1 uses the empirical surface corrections of \cite{Kjeldsen2008}, while AMP Model 2 uses the scaled solar surface corrections of \cite{2012AN....333..914C}.  The YREC model uses the surface corrections of \cite{2014A&A...568A.123B}.}
\label{AMPYREC_Echelle}
\end{figure*}

 For our first optimization runs, we used Scenario B frequencies of Table~\ref{Frequencies} and chose constraints on $\theta$ Cyg luminosity $L = 4.26$ $\pm$ 0.05 L$_\odot$ based on bolometric flux estimate and Hipparcos parallax, log $g$ = 4.2 $\pm$ 0.2 \citep{Erspamer03}, metallicity $-$0.05 $\pm$ 0.15, and radius $R = 1.503$ $\pm$ 0.007 R$_\odot$ \citep{2012A&A...545A...5L}.  Note that these spectroscopic constraints are consistent with, but do not exactly match the final recommended values of Sections \ref{sec:spectroscopy} and \ref{sec:interferometry}.  The AMP (and some preliminary YREC) models were being calculated in parallel with the spectroscopic analyses, and the early asteroseismic results were even used to constrain the log $g$ that was used in the spectroscopic analysis.  The properties of this best-fit model (Model 1) are summarized in Table~\ref{AMPYRECModels}.  Fig.~\ref{AMPYREC_Echelle} shows the \'echelle diagram for this model comparing the observed and calculated frequencies. 

AMP Model 1 has a temperature at the convection-zone base near 320,000 K, exactly right for $\gamma$ Dor $g$-mode pulsations predicted via the convective-blocking mechanism (see Section \ref{sec:gmodepredictions}).  Because we did not find any $g$ modes in the $\theta$ Cyg data, we explored additional models with the final spectroscopic and interferometric constraints summarized in Column 2 of Table \ref{AMPYRECModels}.  The properties of a second AMP model are summarized in Table \ref{AMPYRECModels}.  AMP Model 2 gives an excellent fit to the observed frequencies (see Fig.~\ref{AMPYREC_Echelle}).  Note that the Model 2 \'echelle diagram uses the scaled surface corrections of \cite{2012AN....333..914C} instead of those of \cite{Kjeldsen2008}, improving the match to the high-frequency modes.  However, Model 2 has $T_{\rm eff}$ and radius slightly lower than the spectroscopic constraints, resulting in a low mass and luminosity compared to AMP Model 1 or to the YREC models.  Model 2 has a rather high initial helium mass fraction (0.291), which combined with a lower metallicity (0.0157) compared to Model 1, results in a temperature at the convection-zone base of $\sim$350,000 K, not much higher than for Model 1, despite the lower mass and $T_{\rm eff}$ of Model 2.  Note also that the age of Model 2 is more consistent with that of the best-fit YREC model.

\begin{table*}
\caption{$\theta$ Cyg Frequencies ($\mu$Hz) Identified for Scenario B used in Asteroseismic Modeling Portal}
\begin{center}
\begin{tabular}{ccc} 
\tableline
$l$ = 0 frequency  & $l$ = 1 frequency  & $l$ = 2 frequency \\
\tableline
1086.36  $\pm $  0.15 & 1038.35   $\pm$   0.82   &   996.59  $\pm$  3.82\\
1167.53   $\pm $    0.07  & 1122.87   $\pm$   1.94 & 1083.06 $\pm$   0.25\\
1249.77  $\pm $  0.28   & 1207.90   $\pm$    1.55  & \\
1329.96   $\pm $   0.20   & 1288.13  $\pm$    0.66 & \\
1411.84   $\pm $    0.43  & 1368.96  $\pm$   0.31  & \\
1493.41   $\pm $   0.37   & 1450.85   $\pm$    0.28  & 1405.70 $\pm$   0.87\\
1578.48  $\pm $    0.45   & 1533.79   $\pm$    0.28 & 1487.17  $\pm$   1.49\\
1661.52   $\pm $   0.51   & 1619.81  $\pm$   0.27   &  1573.61  $\pm$   1.39\\
1746.90  $\pm $    0.68   & 1703.47   $\pm$    0.26  & 1658.62  $\pm$   1.33\\
1830.76   $\pm $    0.43   & 1787.82   $\pm$    0.24    & 1743.02  $\pm$  1.18\\
1912.95  $\pm $   0.47   & 1871.74   $\pm$    0.33  & 1826.99  $\pm$   1.05\\
1996.41   $\pm $   0.63   & 1954.68   $\pm$    0.31 & \\
2082.14   $\pm $    0.59  & 2037.49  $\pm$    0.35 & \\
2166.77  $\pm $    0.73   & 2120.73   $\pm$    0.31 &  2079.39  $\pm$   2.68\\
2250.35   $\pm $    0.44   & 2207.91   $\pm$    0.40   &  2160.91  $\pm$   4.22\\
2335.22   $\pm $    0.60   & 2292.26  $\pm$   0.41   &  2243.93  $\pm$   4.1\\
2420.55   $\pm $    0.31   & 2377.88   $\pm$    0.49  & 2326.40  $\pm$   1.51 \\
2507.82  $\pm $    0.89   & 2462.11  $\pm$    0.46  & 2413.11  $\pm$   2.20 \\
2591.20   $\pm $   0.60   & 2547.92  $\pm$    0.58  & 2500.15  $\pm$   2.35\\\
                                          & 2630.50  $\pm$   0.71  &   \\
\tableline
\end{tabular}
\end{center}
\label{Frequencies}
\end{table*}

\begin{table*}
\caption{Observationally-Derived Parameters (Sections \ref{sec:spectroscopy} and \ref{sec:interferometry}), and Properties of AMP and YREC Models}
\begin{center}
\begin{tabular}{lccccc} 
\tableline
      & Observations   & AMP$^{e}$ & AMP$^{e}$ &  YREC Ensemble  & YREC  \\
            &     &    Model 1     & Model 2 &                Average & Best-Fit Model  \\
\tableline
Mass (M$_{\odot}$) &  &   1.39 & 1.26  &     1.346 $\pm$ 0.038 & 1.356\\
Luminosity  (L$_{\odot}$)   &    &  4.215 & 3.350 &  4.114 $\pm$ 0.156 &  4.095\\
$T_{\rm eff}$ (K)  &  6697 $\pm$ 78 &  6753   & 6477 &    6700 $\pm$ 49 & 6700\\
Radius (R$_{\odot}$) & 1.49 $\pm$ 0.03 & 1.503  & 1.457 &    1.507 $\pm$ 0.016 & 1.504\\
log $g$ & 4.23 $\pm$ 0.03 & 4.227 & 4.211 &  4.210 $\pm$ 0.005 & 4.216\\
{[Fe/H]}  &  $-$0.02 $\pm$ 0.06  &  &   &      &  \\
{[M/H]}  &     & 0.028 & $-$0.005 &   $-$0.017 $\pm$ 0.042 & $-$0.035\\
Initial $Y^{a}$   & &   0.276   & 0.291  &  0.272 $\pm$ 0.017 &  0.26475\\
Initial $Z^{b}$  &  &   0.01845  & 0.0157 &   $-$0.0158   & 0.015287 \\
$\alpha$$^{c}$ &   &  1.90  & 1.52  &     1.77 $\pm$ 0.14 & 1.69\\
Age (Gyr)  &  &   0.999  & 1.568 &     1.625 $\pm$ 0.171 & 1.516\\
$T$ CZ$^{d}$ base (K) &  &    320,550   &  354,200 &    & 391,916 \\
$\chi^2$ seismic$^{f}$  & &   9.483  & 8.860 &  & 10.67\\
$\chi^2$ spectroscopic$^{f}$ &  &  0.270 & 2.414 &  & 0.0644\\
\tableline
\end{tabular}
\end{center}
\tablenotetext{a}{$Y$ is mass fraction of helium}
\tablenotetext{b}{$Z$ is mass fraction of elements heavier than H and He}
\tablenotetext{c}{Mixing length/pressure scale height ratio}
\tablenotetext{d}{Envelope convection zone}
\tablenotetext{e}{See \cite{2014ApJS..214...27M} for details}
\tablenotetext{f}{$\chi^2$ minimum of models for seismic and spectroscopic constraints.  See \cite{2014ApJS..214...27M} and text for details.}
\label{AMPYRECModels}
\end{table*}

\section{$\gamma$ Doradus Stars and $g$-mode Predictions}
\label{sec:gmodepredictions}

To determine the predicted $\gamma$\,Dor-like $g$ mode periods for the models of Table \ref{AMPYRECModels}, we calculated corresponding models using the updated Iben evolution code \citep[see][]{2000ApJ...542L..57G}.  Recalculating the models using the Iben code was expedient, since at present we do not have an interface mapping the structure of the AMP model to the \citet{1990ApJ...363..227P} nonadiabatic pulsation code that we use for $g$-mode predictions.   We adjusted the mixing length in the Iben-code formulation to match the the radius at approximately the same age as the AMP or YREC best-fit models.  The Iben models then also approximately matched the luminosity and envelope convection-zone depth of the AMP or YREC models.  The models use OPAL \citep{Iglesias1996} opacities, \citet{2005ApJ...623..585F} low-temperature opacities, and the \cite{gn93} abundance mixture.  Table\,~\ref{IbenModels} gives the properties of the Iben models.  While the Iben model initial masses, compositions, and opacities are the same as in the AMP or YREC models, differences in implementation of mixing-length theory, equation of state, opacity table interpolation, nuclear reaction rates, and fundamental physical constants could be responsible for the small differences in model structure.

\begin{table*}
\caption{Properties of Iben-Code Models and $g$-Mode Predictions}
\begin{center}
\begin{tabular}{lccc} 
\tableline
         & Iben & Iben & Iben    \\
                 &    Model 1     & Model 2 &  Model 3   \\
\tableline
Mass (M$_{\odot}$)   &   1.39 & 1.26  &    1.356   \\
Luminosity  (L$_{\odot}$)       &  4.239 & 3.378 &  4.119  \\
$T_{\rm eff}$ (K)  &    6763   & 6489 &  6712 \\
Radius (R$_{\odot}$)  & 1.503  & 1.457 &  1.504 \\
log $g$ &  4.227 & 4.211 & 4.216 \\
Initial $Y^{a}$   &   0.276   & 0.291 &  0.2648 \\
Initial $Z^{b}$   &   0.01845  & 0.0157 &  0.0153  \\
$\alpha$$^{c}$   &  1.60 & 1.30  &   1.64  \\
Age (Gyr)  &    1.04  &  1.60  &   1.49   \\
$T$ CZ$^{d}$ base (K)   &   319,850   &  355,500  &  393,250  \\
Convective Timescale$^{e}$ at CZ base (days)   &   1.08   &  1.54  &  1.88  \\
Largest $g$-mode growth rate per period & 3.7e-06 & 1.4e-06   & 5.7e-07  \\
$l$ = 1 $g$-mode period range (days) & 0.55 to 1.0 & 0.58 to 1.1 &  0.59 to 1.0 \\
$l$ = 1 $g$-mode frequency range ($\mu$Hz) & 12 to 21 & 11 to 20 &  12 to 20 \\
$l$ = 2 $g$-mode period range (days) & 0.35 to 0.89 & 0.34 to 0.65 & 0.34 to 0.76  \\
$l$ = 2 $g$-mode frequency range ($\mu$Hz) & 13 to 33 & 18 to 34 & 15 to 34 \\

\tableline
\end{tabular}
\end{center}
\tablenotetext{a}{$Y$ is mass fraction of helium}
\tablenotetext{b}{$Z$ is mass fraction of elements heavier than H and He}
\tablenotetext{c}{Mixing length/pressure scale height ratio}
\tablenotetext{d}{Envelope convection zone}
\tablenotetext{e}{Local pressure scale height/local convective velocity}
\label{IbenModels}
\end{table*}


For $\gamma$\,Dor stars, the convective-envelope base temperature that optimizes 
the growth rates and number of unstable $g$ modes is predicted to be about 
300,000 K \citep[see][]{2000ApJ...542L..57G, 2003ApJ...593.1049W}.  For models with convective envelopes 
that are too deep, the radiative damping below the convective envelope quenches 
the pulsation driving; for models with convective envelopes that are too shallow, 
the convective timescale becomes shorter than the $g$-mode pulsation periods, and 
convection can adapt during the pulsation cycle to transport radiation, making the 
convective blocking mechanism ineffective for driving the pulsations.



We calculated the $g$-mode pulsations of the Iben code models using the \citet{1990ApJ...363..227P} non-adiabatic
pulsation code, which also was used by \citet{2000ApJ...542L..57G} and \citet{2003ApJ...593.1049W} to 
investigate the pulsation driving mechanism for $\gamma$\,Dor pulsations and first define
the instability-strip location.  The \citet{1990ApJ...363..227P} code adopts the frozen-convection approximation, which is valid for calculating $g$-mode growth rates, with the driving region at the envelope convection-zone base, only if the convective timescale (defined as the local pressure scale-height divided by the local convective velocity) at the convection-zone base is longer than the pulsation period.  This criterion is met for the best-fit models presented here.  Table \ref{IbenModels} gives the convective timescale at the convective envelope base for each model, and the $g$-mode periods (or alternately, frequencies in $\mu$Hz) for the unstable modes of angular degree $l$=1 and $l$=2.   Table \ref{IbenModels} also gives the maximum growth rate (fractional change in kinetic energy of the mode) per period for each model, which decreases with increasing convection-zone depth because of increased radiative damping in deeper layers.

If $g$ modes were to be detected in $\theta$ Cyg, this star would become the first hybrid $\gamma$ Dor--solar-like oscillator.  However, as discussed in Section \ref{sec:gmodesearch}, $g$ modes have not been detected in the data examined so far.  
It is possible that $\gamma$ Dor modes may be visible in high-resolution spectroscopic observations, but not in photometry.  \cite{2015MNRAS.447.2970B} find for $V = 5.74$ $\delta$ Sct/$\gamma$ Dor hybrid star HD 49434 that some $g$ modes found via high-resolution spectroscopy were not detected in CoRoT photometry, and vice versa.  Another possibility discussed by \citet{2000ApJ...542L..57G} is that shear dissipation from turbulent viscosity near the convection-zone base or in an overshooting region below the convection zone may be comparable to the driving, and may quench the pulsations.  The predicted $g$-mode growth rates of $\sim$10$^{-6}$ per period are smaller than typical $\delta$ Sct $p$-mode growth rates of  $\sim$10$^{-3}$ per period.   The models presented here do not take into account diffusive settling, radiative levitation, or changes in abundance mixture that could affect the convection zone depth and $g$-mode driving.  $\theta$ Cyg may therefore be important for furthering our understanding of the role of stellar abundances, diffusive settling, and turbulent convection on stellar structure and asteroseismology.


\section{Search for $g$ modes in $\theta$ Cyg Data}
\label{sec:gmodesearch}

\begin{figure}
\center{\includegraphics[width=1.0\columnwidth]{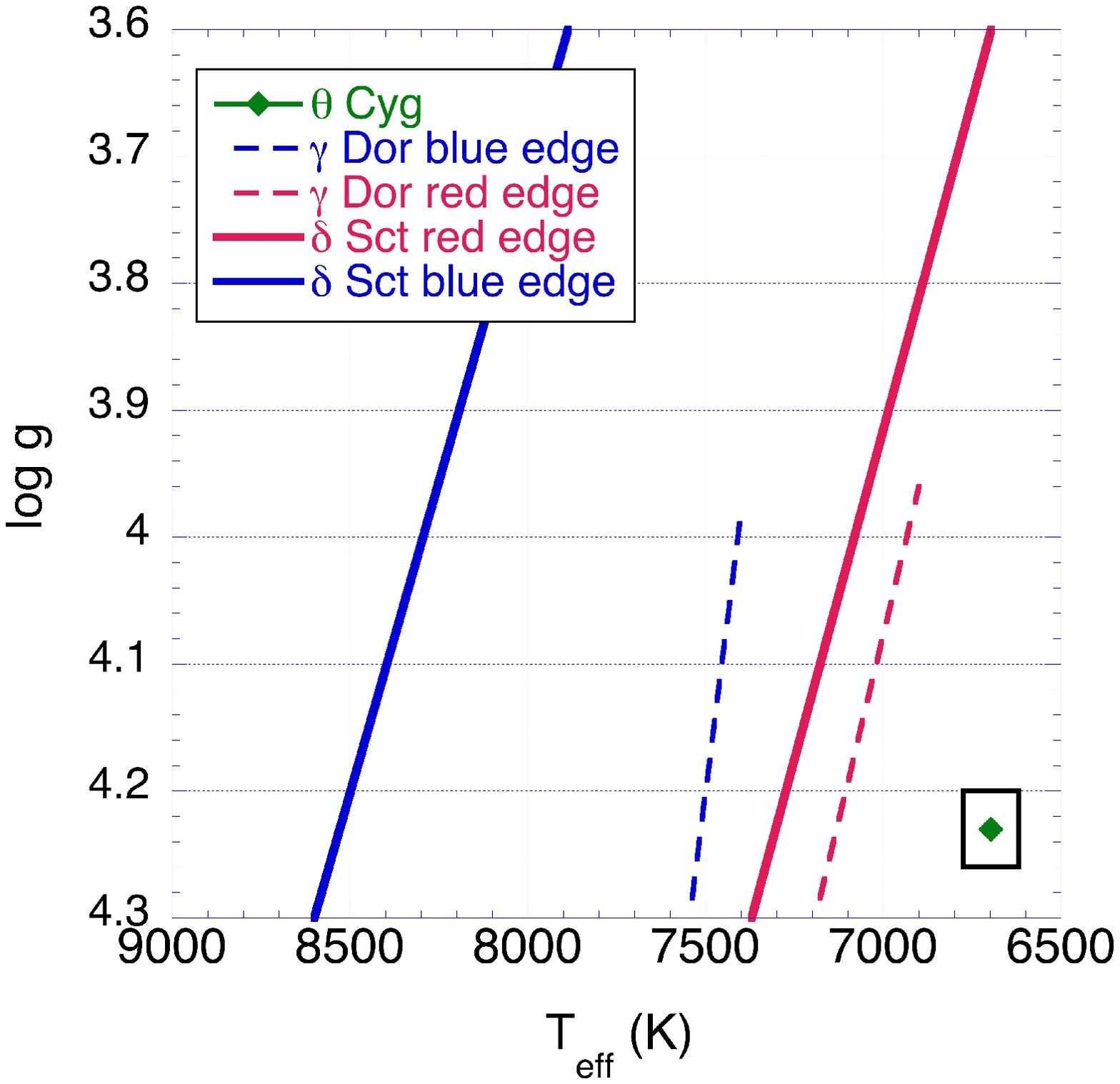}}
\caption{HR~diagram location of $\theta$ Cyg relative to $\gamma$ Dor and $\delta$ Sct instability strip edges established from ground-based observations.  In this figure $\theta$ Cyg is located at 6697 $\pm$ 78 K, and log $g$ = 4.23 $\pm$ 0.03 (see Section \ref{sec:spectroscopy}).} 
\label{thetaCygISHRD}
\end{figure}

Figure~\ref{thetaCygISHRD} shows the location of $\theta$ Cyg relative to the instability strip locations established from ground-based discoveries of $\gamma$ Dor and $\delta$ Sct stars \citep[see][and references therein]{2011A&A...534A.125U}.  The temperature used for $\theta$ Cyg's location in this figure is 6697 $\pm$ 78 K based on the spectroscopic observations summarized in Section \ref{sec:spectroscopy}.
$\theta$ Cyg's log $g$ and effective temperature in this figure places it to the right of the red edge of the $\gamma$ Dor instability strip established from pre-{\it Kepler} ground-based observations.  Taking into account more generous uncertainties on effective temperature and surface gravity, $\theta$ Cyg could be just at the edge of the instability strip.   $\gamma$ Dor candidates have been discovered in the {\it Kepler} data that appear to lie beyond this $\gamma$ Dor red edge based on {\it Kepler} Input Catalog parameters \citep[see, e.g.][]{2011A&A...534A.125U, 2015arXiv150200175G}.  However, the purer sample of {\it Kepler} $\gamma$ Dor stars with log $g$ and $T_{\rm eff}$ established from high-resolution spectroscopy \citep{2015ApJS..218...27V} does fall within the $\gamma$ Dor instability strip established from theory \citep{2013MNRAS.429.2500B}.  See, in addition, \cite{2015MNRAS.450.2764N} and \cite{2013A&A...556A..52T}, who also do not show $\gamma$ Dor stars beyond this red edge. The theoretically derived instability regions of \cite{2013MNRAS.429.2500B} and \citet{2005A&A...435..927D} including time-dependent convection show the red edge extending at log $g$ = 4.2 to $\sim$6760 K, placing $\theta$ Cyg just at the red edge.

In contrast to  stochastically excited solar-like oscillations, $g$-mode pulsations excited by the convective blocking mechanism are known to be coherent, resulting in sharp peaks in the Fourier spectrum, with line widths defined by the duration of the observations. 
Attributing low-frequency signals to $g$ modes requires caution, as other phenomena such as granulation, spots and instrumental effects occur at similar time scales.  However it is possible to distinguish between these signatures. The granulation background noise, for example, as observed in many solar-type stars and red giants \citep[e.g.,][]{2011ApJ...741..119M} but also in $\delta$ Scuti stars \citep[e.g.,][]{2010ApJ...711L..35K,2011ApJ...741..119M}, has a distinct signature which can be described as the sum of power laws with  decreasing amplitude as a function of increasing frequency \citep[e.g.,][]{2010ApJ...711L..35K}. 
Long-lived stellar spots, on the other hand, which follow the rotation often result in a single peak; however, if latitudinal differential rotation occurs, and/or the spot sizes and lifetimes change, as observed in the Sun, spots can produce a peak with a multiplet structure \citep{2009A&A...506..245M, 2010A&A...518A..53M, 2011A&A...530A..97B,2014A&A...572A..34G} which can be misinterpreted as $g$ modes. In the case of stellar activity, the temporal variability allows to draw a conclusion. 
Instrumental effects are not easy to identify; however, in the present case, we can compare the light curve of $\theta$ Cyg with that of other stars, observed during the same quarters and we can also exclude very long periods. Also, contamination by 
background stars needs to be taken into account, especially in the present case, as the collected light is spread over 1600 pixels on the detector.

Visual inspection of the processed light curve (Section \ref{sec:detection}, Fig.~\ref{Q6Q8LightCurve}) indicates that we might observe rotational modulation due to spots, as discussed by \citet{2011MNRAS.415.3531B}. In the Fourier spectrum, we find a peak at 0.159 d$^{-1}$ (1.840 $\mu$Hz), which translates into a period of 6.29 days.  If this were a rotational period, using a radius $R=1.5$ R$_{\odot}$  and $v \sin{i}$ = $3.4 \pm 0.4$ km\,s$^{-1}$ (Section \ref{sec:spectroscopy}), the rotational 
velocity would be 12 km\,s$^{-1}$ and the inclination angle would be 16 $\pm$ 2 degrees. The frequency at 0.159 c d$^{-1}$  is present in both quarters; however at the end of Q8 the amplitude at this frequency starts to diminish, a temporal variability consistent with a changing activity cycle.  Rotational frequencies may also be distinguished from $g$-mode frequencies if the modes behave linearly \citep[see, e.g.,][]{2013A&A...551A..12T}, as the rotational frequency would occur with multiple harmonics, whereas the $g$-mode frequency would not. 

\begin{figure*}
\includegraphics[width=\textwidth]{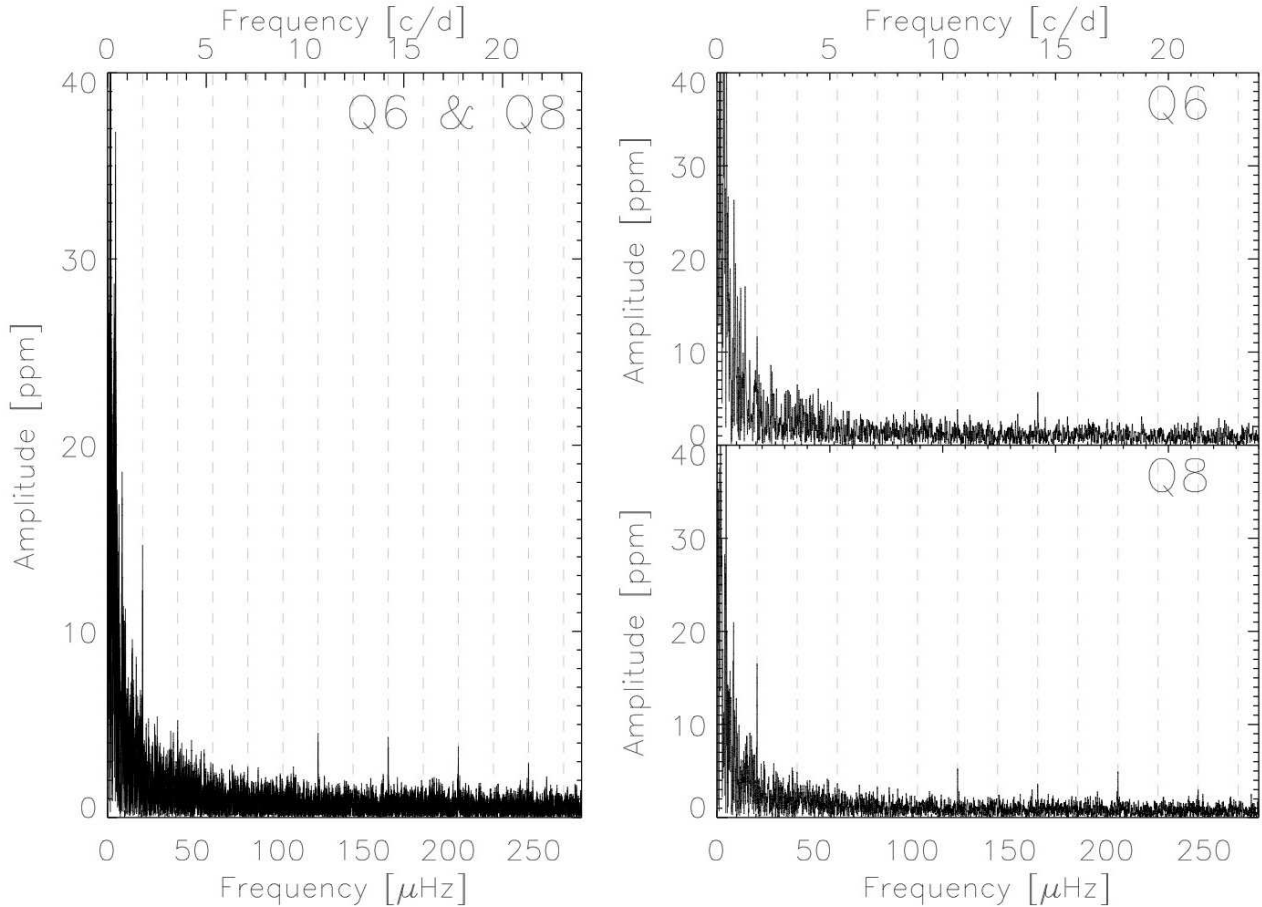}
\caption{Fourier spectra of the individual quarters and the combined data, with the vertical dashed gray lines indicating the orbital frequency of the background binary and its harmonics.  One remaining frequency at $\sim$ 1.7 d$^{-1}$ probably is attributable to a background binary (see text and Fig. \ref{foldedlightcurve}).} 
\label{Antoci_Q6Q8}
\end{figure*}
\begin{figure*}
\includegraphics[width=\textwidth]{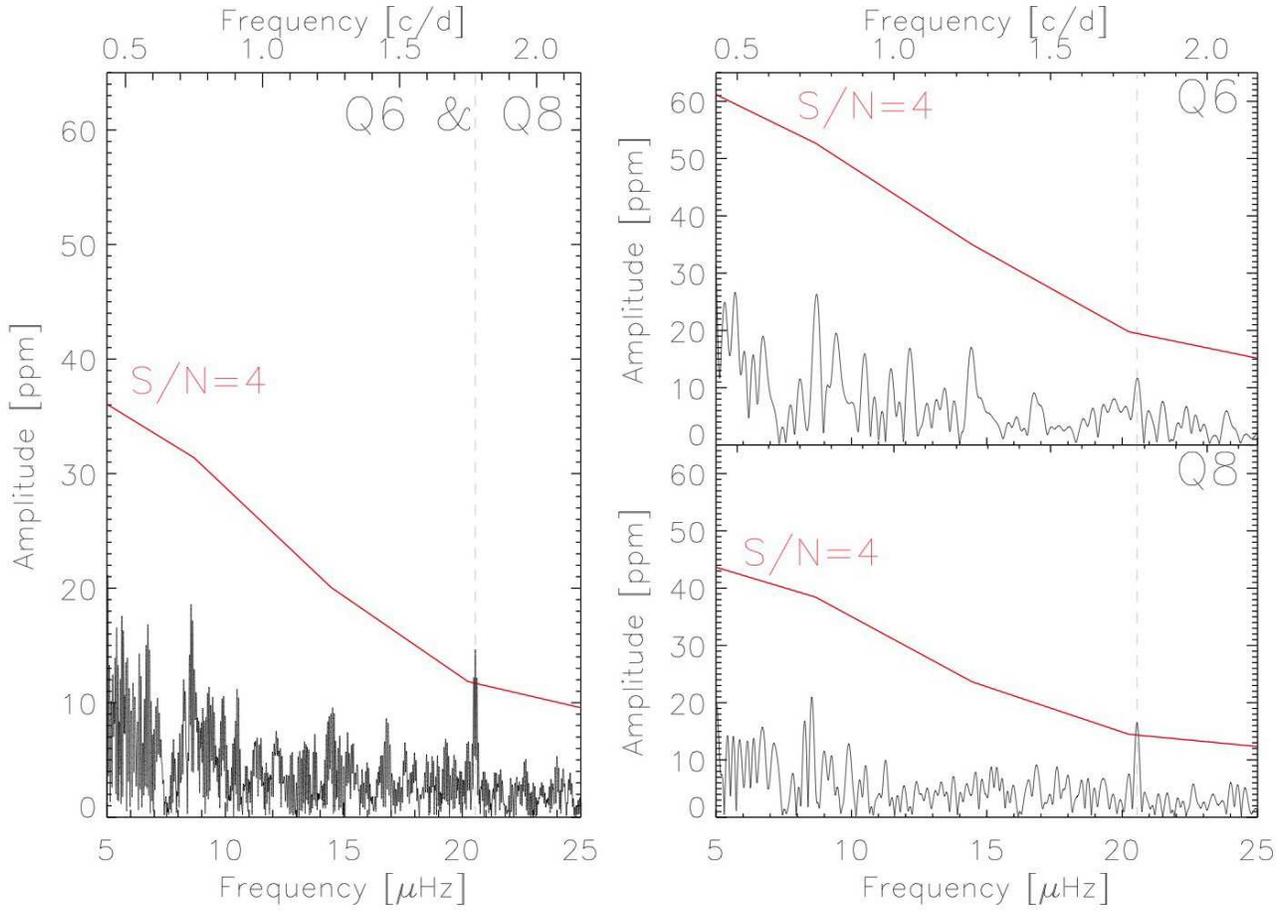}
\caption{Zoom-in of the region from 5 to 25 $\mu$Hz (0.5 to 2.5 d$^{-1}$ ).  The red line indicates the S/N (signal-to-noise) = 4.0 significance criterion calculated using Period04 \citep{2005CoAst.146...53L}.} 
\label{Antoci_Q6Q8_zoom}
\end{figure*}
\begin{figure*}
\mbox{
\includegraphics[scale=0.3]{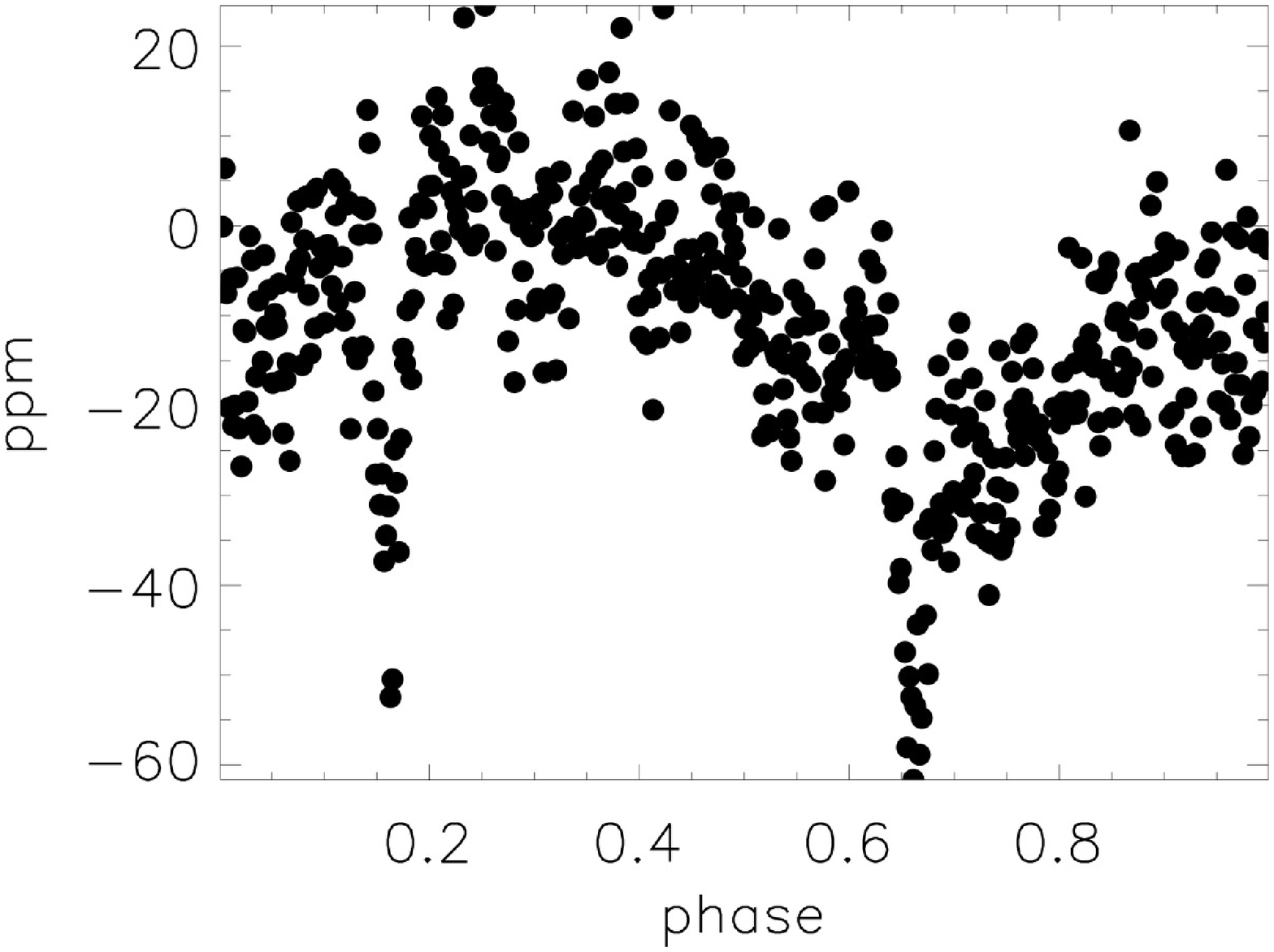}
\includegraphics[scale=0.3]{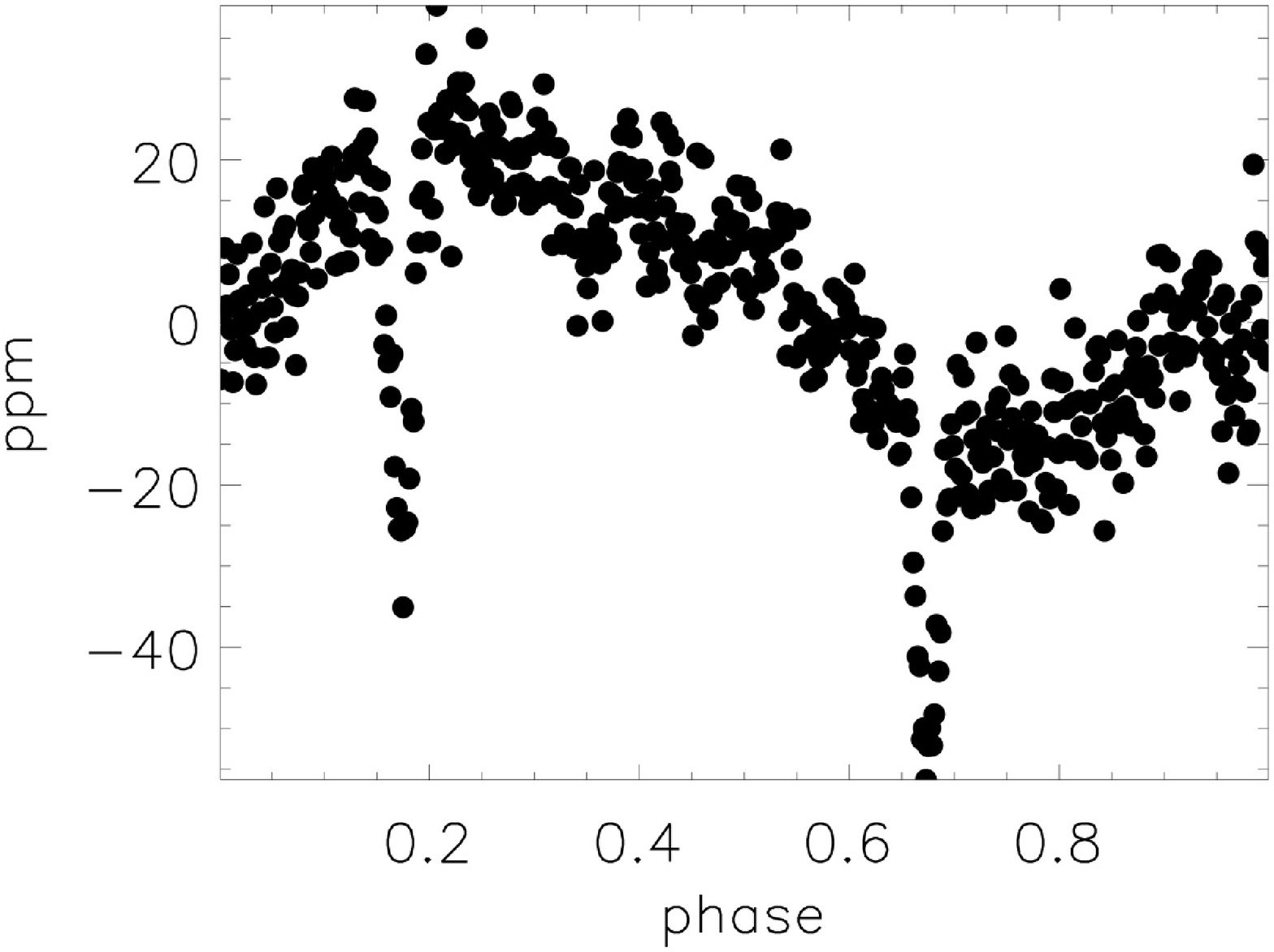}
}
\caption{Q6 (left) and Q8 (right) light curve folded at 1.77627 d$^{-1}$ ($\sim$20.56 $\mu$Hz), showing that a faint background binary is the likely explanation for this frequency in the power spectrum.} 
\label{foldedlightcurve}
\end{figure*}

To search for $g$ modes, we analyzed the short-cadence Q6 and Q8 data separately, and then in combination.
Figure \ref{Antoci_Q6Q8} shows the amplitude spectrum, and Figure \ref{Antoci_Q6Q8_zoom} shows a zoom-in of this spectrum for frequencies from 5 to 25 $\mu$Hz (0.43 to 2.16 d$^{-1}$).  We find one significant peak at 20.56 $\mu$Hz (1.7763 d$^{-1}$), which is a good candidate for a $g$ mode, but one peak alone is usually not enough to claim the detection of such pulsation modes.  From {\it Kepler} observations we know that $\gamma$ Dor stars as well as $\gamma$ Dor/$\delta$ Sct hybrids usually show more than one $g$ mode excited \citep{2013A&A...556A..52T}.  In the present case however we can definitely exclude this frequency from being a $g$ mode, because the binned phase plot clearly shows the signature of a binary system, which is around 10 magnitudes fainter than $\theta$ Cyg.   Figure~\ref{foldedlightcurve} shows the binned phase plot folded by 1.7763 d$^{-1}$ for the different quarters.   Figures \ref{Antoci_Q6Q8} and \ref{Antoci_Q6Q8_zoom} show with vertical dashed gray lines the harmonics of this frequency.

We have not established whether the binary signal is related to the $\theta$ Cyg system.  Identifying the source of the binary signal and its relationship to $\theta$ Cyg would require considerable work given the faintness of the source.   One could investigate whether the signal is more prominent in the point-spread function by comparing the Fourier transform of data sets with different extraction masks, covering different parts of the point-spread function; if the signal is associated with $\theta$ Cyg, additional radial velocity measurements may also be required.

A question of interest is the effect of the binary signal on the light curve on the derived $p$-mode oscillation properties.   In order to affect the signal in the 1000-3000 $\mu$Hz region of the $p$-mode spectrum, the signal would need to be approximately the 50th harmonic of the 1.7763 d$^{-1}$ binary frequency.  Such high harmonics would not be visible, especially considering that the base frequency is barely significant, as shown in Fig. \ref{Antoci_Q6Q8_zoom}. The $p$-mode amplitudes, converting from ppm$^2$/$\mu$Hz to ppm, are approximately 30 to 100 ppm, while the binary signal harmonics near 250 $\mu$Hz already have amplitudes as low as $\sim$2 ppm, and will become even smaller at higher frequencies.

In addition, simulations have been performed \citep[see supplementary on-line information for][]{2011Natur.477..570A} in the context of KIC 7548479 for artificial data containing coherent non-stochastic signals (binary and $g$ modes) and non-coherent solar-like oscillations, to understand whether prewhitening the coherent signals influences the non-coherent ones.  It was found that prewhitening the coherent signals does not affect the solar-like oscillations. 

It is interesting that the eclipsing binary orbital frequency is close to one-fourth of the large separation (4 x 20.56 $\mu$Hz = 82.24 $\mu$Hz $\simeq$ 83.9 $\mu$Hz).   A single star orbiting $\theta$ Cyg at this period would have an orbital semi-major axis of $\sim$3.2 R$_{\odot}$, a little over twice $\theta$ Cyg's radius.   Another possibility is that a binary system with this orbital period is associated with $\theta$ Cyg (see discussion of $\theta$ Cyg B in the Appendix).  In either case, it could be considered whether tidal effects could have some effect on the $p$-mode spacing.  Tidal effects have been shown to drive modes separated by the orbital frequency in the so-called heartbeat stars \citep{2012ApJ...753...86T,2015EPJWC.10104007H}, but the modes driven are generally in the $g$-mode range.  Additional shorter periods are also found in some heartbeat stars, and at least one heartbeat star, KIC 4544587, has some $\delta$ Sct $p$-modes separated by multiples of the orbital frequency \citep{2013MNRAS.434..925H} .  However, if there were tidal forcing involved, we would expect the pulsation periods to be exact multiple integers of the orbital period, which is not the case for $\theta$ Cyg.  Furthermore, such modes would be expected to be coherent, unlike the stochastically excited $p$ modes observed for $\theta$ Cyg.  Therefore, we consider an association between the binary frequency and the $\theta$ Cyg $p$ modes to be unlikely.  

The binary signal has very small amplitude, barely above the signal-to-noise criterion of 4, which means that if $g$ modes were present they should be visible in the spectrum of Figs.~\ref{Antoci_Q6Q8} and \ref{Antoci_Q6Q8_zoom}.  We have done some tests with the long-cadence data, prewhitening for the binary harmonic, and find no other significant long-period modes.

\section{Conclusions and Motivation for Continued Study of $\theta$ Cyg}
\label{sec:conclusions}

We have analyzed Quarters 6 and 8 of {\it Kepler} $\theta$ Cyg data, finding solar-like $p$-modes, and not finding $\gamma$ Dor gravity modes that were initially expected given $\theta$ Cyg's spectral tye.   We have obtained new ground-based spectroscopic and interferometric observations and updated the observational constraints.  Stellar models of $\theta$ Cyg that fit the $p$-mode frequencies and spectroscopic and interferometric constraints on $R$, $T_\mathrm{eff}$, log $g$, and [M/H] are predicted to show $g$-mode pulsations driven by the convective-blocking mechanism, according to nonadiabatic pulsation models.  However, analysis of the light curves did not reveal any $g$ modes.

Reprocessed {\it Kepler} observations of $\theta$ Cyg for Quarters 12 through 17 including the pipeline corrections will be available in late 2016.  We intend to examine the pixel-by-pixel data to remove the background binary if possible.  As noted by \cite{2013A&A...556A..52T} in analysis of their sample of 69 $\gamma$ Dor stars, use of the pixel data eliminated many spurious low frequencies detected using the standard pre-processed light curves.   Analyses of a longer time series may reduce noise due to granulation, and more definitively rule out the presence of $g$ modes or identify features in the light curve resulting from rotation and stellar activity.  The {\it Kepler} observations of $\theta$ Cyg, in conjunction with studies of many other A-F stars observed by {\it Kepler} and CoRoT, will be key to understanding the puzzles of $\gamma$ Dor/$\delta$ Sct hybrids and pulsating variables that appear to lie outside of instability regions expected from theoretical models, and to test stellar model physics and possible alternative pulsation driving mechanisms.

Attempting to find $g$ modes in $\theta$ Cyg and other mid-F spectral type stars is worthwhile, as $g$ modes are more sensitive to the stellar interior near the convective core boundary than are $p$ modes.   Seismic measurements of convective core size and shape, and the structure of the overshooting region will help reduce uncertainties in stellar ages and understand the roles of penetrative overshooting and diffusive mixing.   Progress has already been made in this area for {\it Kepler} slowly-pulsating B stars that are $g$ mode pulsators by, e.g., \cite{2015A&A...580A..27M}, who used the spacings of 19 consecutive $g$ modes in KIC 9526294 to distinguish between models using exponentially decaying vs. a step-function overshooting prescription, and diagnose the need for additional diffusive mixing.  However, note that progress is also being made studying convective cores using $p$ modes in low-mass stars \citep[see, e.g.][]{2016arXiv160302332D}, as the molecular weight gradient outside the convective core introduces a discontinuity in sound-speed profile that is diagnosable with $p$ modes.

The core size and mode frequencies are also affected by rotation that is likely to be more rapid in the core than in the envelope.  \cite{2015ApJS..218...27V} discuss $g$-mode periods and spacings for a sample of 67 $\gamma$ Dor stars observed by {\it Kepler},  and find correlations between $v~\sin i$, $T_{\rm eff}$, period spacing values, and dominant periods.  van Reeth et al. (2016, submitted), discuss a method for mode identification of high-order $g$ modes from the period spacing patterns for $\gamma$ Dor stars, allowing to deduce rotation frequency near the core.  \cite{2015EPJWC.10101005B} discuss using period \'echelle diagrams for {\it Kepler} $\gamma$ Dor stars  to measure period spacings and identify rotationally split multiplets with $l$ = 1 and $l$ = 2. \cite{2015MNRAS.454.1792K} study KIC 10080943, two hybrid $\delta$ Sct/$\gamma$ Dor stars in a non-eclipsing spectroscopic binary, and are able to use rotational splitting to estimate core rotation rates.

Because $\theta$ Cyg is nearby and bright, and data can be obtained with excellent precision, it is also a worthwhile target for continued long time-series ground- or space-based photometric or spectroscopic observations.  With an even longer time series of data (obtainable by a follow-on to the {\it Kepler} mission), there is the possibility to study rotational splitting and differential rotation, infer convection zone depth directly from oscillation frequency inversions, measure sin~$i$ directly from amplitude differences of rotationally split modes, and investigate possible magnetic activity cycles.  

\acknowledgments

We are grateful to the {\it Kepler} Guest Observer program for observing $\theta$ Cyg with a custom aperture.  We thank the referee for helpful comments and suggestions.  J.A.G. acknowledges support from {\it Kepler} Guest Observer grant KEPLER08-0013, NASA Astrophysics Theory Program grant 12-ATP12-0130, and the KITP Asteroseismology Institute at U.C. Santa Barbara in December 2011.   G.H. acknowledges support from the Austrian FWF Project P21205-N16.  R.A.G., G.R.D., and K.U. have received funding from the European Community's Seventh Framework Program (FP7/2007-2013) under grant agreement no. 269194.  K.U. acknowledges support by the Spanish National Plan of R\&D for 2010, project AYA2010-17803.  S.B. acknowledges support from NSF grants AST-1514676 and AST-1105930,  and NASA grants NNX16AI09G and NNX13AE70G.  P.I.P. is a Postdoctoral Fellow of The Research Foundation -- Flanders (FWO), Belgium, and he also acknowledges funding from the Belgian Science Policy Office (BELSPO, C90309: CoRoT Data Exploitation).  S.H. acknowledges funding from the European Research Council under the European Community's Seventh Framework Programme (FP7/2007-2013)/ ERC grant agreement number 338251 (StellarAges).  J.M-\.Z. acknowledges the Polish Ministry grant No. NCN 2014/13/B/ST9/00902.  Funding for the Stellar Astrophysics Centre is provided by the Danish National Research Foundation (Grant DNRF106).  The research is supported by the ASTERISK project (ASTERoseismic Investigations with SONG and {\it Kepler}) funded by the European Research Council (Grant agreement no.: 267864). S.G.S acknowledges the support from the Fundação para a Ciência e Tecnologia (Portugal) in the form of the grant SFRH/BPD/47611/2008. R.A.G. received funding from the European Community's Seventh Framework Programme (FP7/2007-2013) under grant agreement No. 312844 (SPACEINN).  B.M. and R.A.G. received funding from the ANR (Agence Nationale de la Recherche, France) program IDEE (n ANR-12-BS05-0008) ``Interaction Des \'Etoiles et des Exoplan\`etes''. R.A.G., G.R.D., and D.S. acknowledge support from the CNES. S.M. acknowledges support from the NASA grant NNX12AE17G. D.W.L. acknowledges partial support from the {\it Kepler} mission under NASA Cooperative Agreements NNX11AB99A and NNX13AB58A with the Smithsonian Astrophysical Observatory.


\appendix
\label{sec:background}

\section{The $\theta$ Cygni System}

\begin{figure*}
\includegraphics[width=\columnwidth]{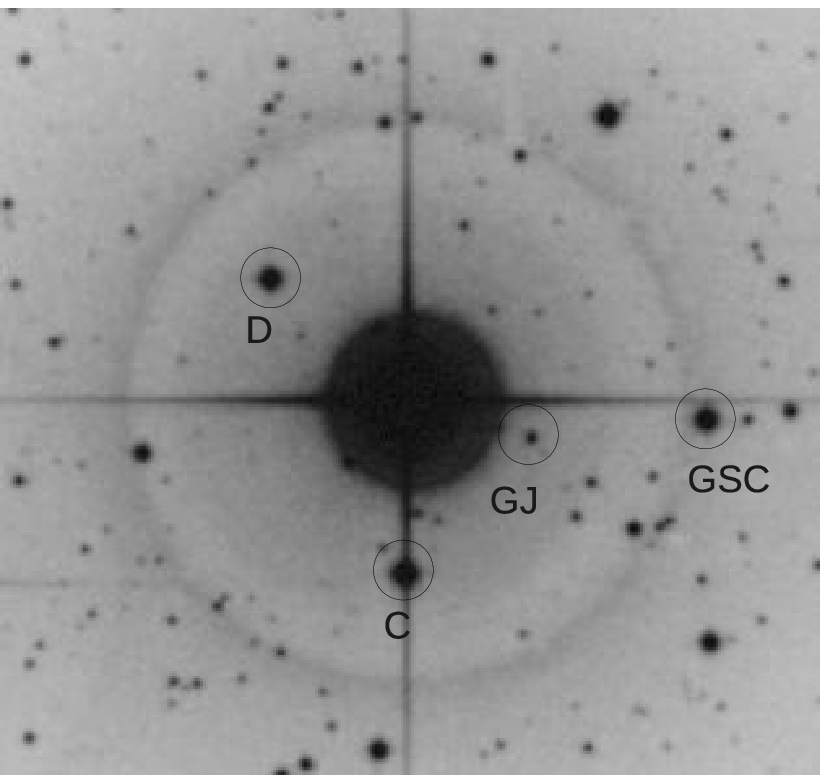}
\caption{Digitized Sky Survey image showing the field around $\theta$ Cyg AB. The scale of this figure is approximately 5$\frac{1}{4}$ arc minutes wide.  The stars discussed in the text are highlighted. GJ is GJ765B and GSC is GSC 03564-00642.}
\label{field}
\end{figure*}

The field around $\theta$ Cyg (Fig.~\ref{field}) has been examined to identify
any stars which might be part of the system. Selected stars are now discussed.
For the other stars within 2{\arcmin} there is insufficient evidence to suggest
that they are companions of $\theta$ Cyg.

\subsection{$\theta$ Cyg A}

There is a vast literature on $\theta$ Cyg A, that is summarized in
Table~\ref{literature}. This review of the literature prior to the {\it Kepler}
observations shows that $\theta$ Cyg A is a normal slowly-rotating
solar-composition F3V-type star \citep{2003AJ....126.2048G} with $T_\mathrm{eff}$ around 6700$\pm$100~K and
$\log g$ around 4.3$\pm$0.1 dex.

\begin{table*}
\begin{center}
\caption{A summary of parameter determinations of $\theta$ Cyg in the literature.
While not all referenced values are new and independent determinations, the compilation
does give an indication of the range of values previously found. At the bottom of table
the averages and standard deviations are given, in order to indicate the typical scatter in results.}
\begin{tabular}{ccccc} \hline
$T_\mathrm{eff}$&$\log g$&$[Fe/H]$ & Reference \\ \hline
6700 &         &      &  {\citet{Bohm-Vitense78}} \\
7000 &   4.27  & 0.07 &  {\citet{Philip80}} \\ 
6545 &   4.40  & -0.21&  {\citet{Thevenin86}} \\
6632 &   4.40  &  0.10&  {\citet{1986ApJ...309..762B}} \\
6840 &         &      &  {\citet{Malagnini90}} \\
6770 &   4.41  &      &  {\citet{1991MNRAS.252..329A}} \\
6713 &         &      &  {\citet{Blackwell94}} \\
6725 &   4.35  & 0.01 &  {\citet{1995BICDS..47...13M}} \\
6462 &         &  0.04&  {\citet{1966ApJ...143..336M}} \\
6550 &   4.4   & 0.00 &  {\citet{1998BICDS..49.....T}} \\
6672 &         &      &  {\citet{Blackwell98}} \\
6666 &         &      &  {\citet{diBenedetto98}} \\
6760 &   4.24  &      &  {\citet{Prieto99}} \\
6700 &   4.30  &  0.01&  {\citet{2000ApJ...530..939C}} \\
6640 &         &-0.02 &  {\citet{Taylor03}} \\
6745 &   4.21  &-0.03 &  {\citet{Erspamer03}} \\
6704 &   4.35  &-0.02 &  {\citet{LeBorgne03}} \\
6747 &   4.21  &-0.04 &  {\citet{2003AJ....126.2048G}} \\
6594 &   4.04  &-0.03 &  {\citet{2005ApJS..159..141V}} \\
6810 &         & 0.1  &  {\citet{2005AstL...31..388R}} \\
     &   4.20  &      &  {\citet{2007ApJS..168..297T}} \\
6650 &         &-0.04 &  {\citet{Holmberg09}} \\
\hline\hline
6696       &    4.29     &      0.00 \\
$\pm$ 115  &  $\pm$0.11  & $\pm$ 0.08 \\ \hline
\end{tabular}
\label{literature}
\end{center}
\end{table*}

\subsection{$\theta$ Cyg B}

The close companion $\theta$ Cyg B (KIC 11918644; 2MASS 19362771+5013419) is
listed in the {\it Washington Visual Double Star Catalog} (WDS)
\citep{2001AJ....122.3466M} as a magnitude 12.9 star at 3.6{\arcsec} and PA 44{\degr}
in 1889. The orbital motion was discussed by \cite{2009A&A...506.1469D}, who
give a projected separation 46.5\,AU, a minimum period of roughly 230 years and a
mass from evolutionary codes of 0.35\,$M_{\sun}$.

Using the $H$ and $K$ contrasts given in \cite{2009A&A...506.1469D}, we estimate
that $H \sim 8.3$ and $K \sim 8.0$. Using $V \sim 12.9$, an approximate
bolometric flux of $F_{\rm bol} \sim 1.0^{-12}$~W\,m$^{-2}$ was obtained.
Using the IRFM \citep{1977MNRAS.180..177B}, we estimate that $T_\mathrm{eff}$ = $3000\sim3500$~K and an
angular diameter of $\sim0.18$~mas. Using Hipparcos distance, we get $\log L/ L_{\sun} \sim -2.0$, $M_{\rm bol} \sim 9.7$ and $R \sim 0.36\,R_{\sun}$.

\begin{figure}
\includegraphics[width=\columnwidth]{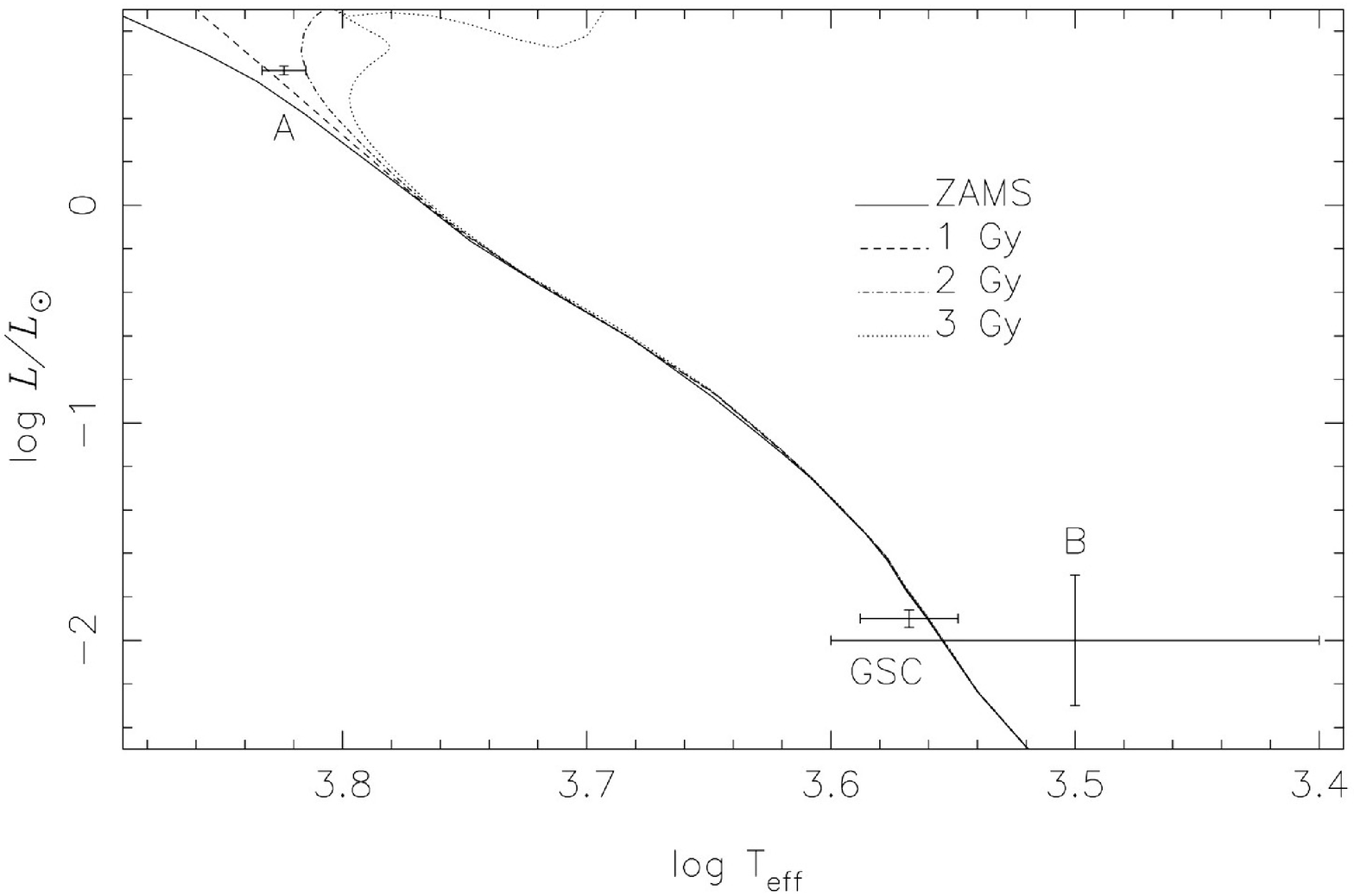}
\caption{Hertzsprung-Russell diagram for $\theta$ Cyg system, showing components A and B, plus
the common proper-motion companion GSC 03564-00642.  Models are from \cite{2008A&A...482..883M}.}
\label{HR-Diagram}
\end{figure}

The approximate position of $\theta$ Cyg B in the HR diagram is shown in Fig.~\ref{HR-Diagram}.

In Section~\ref{sec:gmodesearch} we identified a potential short-period binary within the {\it
Kepler} mask. If this star is the binary and has equal components, then the
individual stars have mass of $\sim0.18\,M_{\sun}$ and radii of $\sim
0.25\,R_{\sun}$. The individual luminosities will be 0.3~dex lower, placing
them closer to the isochrone in Fig.~\ref{HR-Diagram}. 

\subsection{$\theta$ Cyg C}

The {\it Bright Star Catalogue} states that WDS 19364+5013AC (KIC 11918629) is a
mag. 11.6 optical companion at 29.9{\arcsec} and PA 186{\degr} in 1852. The
current separation of $\sim$1{\arcmin} supports this conclusion.

\subsection{$\theta$ Cyg D}

WDS 19364+5013AD (KIC 11918668) is a mag. 12.5 $T_\mathrm{eff}$ = 6800~K star at
82.1{\arcsec} and PA 40{\degr} in 1923. With a current separation of
1.17{\arcmin} and PA 50{\degr} this is an optical companion.

\subsection{GJ 765B}

With similar proper motions (\cite{2005AJ....129.1483L}), the star GJ 765B
(2MASS 19362286+5013034; KIC 11918614) could be a common proper motion
companion. Optical photometry ($V \sim 13.03$) and 2MASS suggest that this could
be a hot star (A or B-type). The estimated $\log L/L_{\sun} \sim -2.0$  and $M_{\rm bol}
\sim 9.9$ are inconsistent for a main-sequence star, but not a subdwarf.

Alternatively, this star might actually be 2MASS19362147+5012599 (KIC 11918601),
but for the same $V$ magnitude this would also be a hot star with low luminosity
($\log L/L_{\sun} \sim -2.4$  and $M_{\rm bol} \sim 10.7$).

Further observations are required to confirm the nature of these stars, in order
to determine whether or not this is a common proper motion companion.

\subsection{GSC 03564-00642}

\cite{1924AJ.....35..180Y} suggested that a faint companion (GSC 03564-00642,
2MASS J19361440+5013096; KIC 11918550) 2{\arcmin} west of $\theta$~Cyg~A was
physical. \cite{1980PASP...92..345B} confirmed that the spectral type, M2/3 is
consistent with this suggestion. The proper motion is slightly different from
$\theta$~Cyg~A, but not totally inconsistent with this suggestion considering
the range of values in the various catalogues.

Available broad-band photometry, suggests $T_\mathrm{eff} \sim 3700$~K,
$F_\mathrm{bol} \sim 1.2^{-12}$~W\,m$^{-2}$ and angular diameter
$0.14\pm0.03$~mas. Using the Hipparcos distance to $\theta$~Cyg~A, we get $\log L/L_{\sun} = -1.90\pm0.04$,
$M_{\rm bol} = 9.50\pm0.11$ and $R = 0.28\pm0.06$~R$_{\sun}$. The position in HR
Diagram is also shown in Fig.~\ref{HR-Diagram}, and the star appears to be part of the
$\theta$ Cygni system.





\bibliographystyle{aastex6}



\end{document}